\documentclass[aps,showpacs,twocolumn,prd,superscriptaddress,nofootinbib]{revtex4}

\usepackage{amsmath}
\usepackage{graphicx}
\usepackage{epstopdf}
\usepackage{hyperref}
\usepackage{url}
\usepackage{color}
\usepackage{amssymb}
\usepackage{wasysym}
\usepackage{mathtools}
\usepackage[letterpaper,margin=0.75in]{geometry}

\newcommand{\Msun}{\>{\rm M_{\odot}}}
\newcommand{\beq}{\begin{equation}}
\newcommand{\eeq}{\end{equation}}
\newcommand{\bea}{\begin{eqnarray}}
\newcommand{\eea}{\end{eqnarray}}

\def\leq{\raise 0.4ex\hbox{$<$}\kern -0.8em\lower 0.62ex\hbox{$-$}}
\def\geq{\raise 0.4ex\hbox{$>$}\kern -0.7em\lower 0.62ex\hbox{$-$}}
\def\lsim{\raise 0.4ex\hbox{$<$}\kern -0.8em\lower 0.62ex\hbox{$\sim$}}
\def\gsim{\raise 0.4ex\hbox{$>$}\kern -0.7em\lower 0.62ex\hbox{$\sim$}}
\def\appropto{\raise 0.4ex\hbox{$\propto$}\kern -0.7em\lower 0.62ex\hbox{$\sim$}}
\def\pm{\,\raise 0.4ex\hbox{$+$}\kern -0.8em\lower 0.62ex\hbox{$-$}\,}

\begin{document}

\title{Detecting gravitational waves from highly eccentric compact binaries}
\author{Kai Sheng Tai}
\affiliation{Department of Physics, Princeton University, Princeton, NJ 08544}
\author{Sean T. McWilliams}
\email{sean.mcwilliams@mail.wvu.edu}
\affiliation{Department of Physics, Princeton University, Princeton, NJ 08544}
\affiliation{Department of Physics and Astronomy, West Virginia University, Morgantown, WV 26503}
\author{Frans Pretorius}
\affiliation{Department of Physics, Princeton University, Princeton, NJ 08544}

\date{\today}


\begin{abstract}
In dense stellar regions, highly eccentric binaries of black holes and neutron stars can form through various n-body interactions. Such a binary could emit a significant fraction of its binding energy in a sequence of largely isolated gravitational wave bursts prior to merger. Given expected black hole and neutron star masses, many such systems will emit these repeated bursts at frequencies within the sensitive band of contemporary ground-based gravitational wave detectors. Unfortunately, existing gravitational wave searches are ill-suited to detect these signals. In this work, we adapt a ``power stacking'' method to the detection of gravitational wave signals from highly eccentric binaries. We implement this method as an extension of the Q-transform, a projection onto a multiresolution basis of windowed complex exponentials that has previously been used to analyze data from the network of LIGO/Virgo detectors. Our method searches for excess power over an ensemble of time-frequency tiles. We characterize the performance of our method using Monte Carlo experiments with signals injected in simulated detector noise. Our results indicate that the power stacking method achieves substantially better sensitivity to eccentric binary signals than existing localized burst searches.
\end{abstract}


\maketitle

\section{Introduction}
\label{sec:intro}
Stellar mass compact object binary coalescences are among the most promising sources of gravitational wave (GW) emission that are hoped to be detected by the next generation of ground-based GW observatories, including LIGO~\cite{abramovici1992ligo},
VIRGO~\cite{caron1997virgo}, GEO600~\cite{willke2002geo}, and KAGRA(LCGT)~\cite{Ohashi:2011zz}. Relevant compact objects are black holes (BH) and neutron stars (NS), and when the latter are involved correlated electromagnetic and neutrino emission could occur. Though neutrinos will not be detectable except for (exceedingly rare) events within our galactic neigbourhood, a broad class of putative electromagnetic counterparts could be observed to distances where GW emission will be within reach of the detectors. These have been suggested to include short gamma ray bursts, kilo- or macro-nova (radio to UV emission on timescales of a day to a week from radioactive decay of ejected material~\cite{Metzger2012,Piran2012,2013arXiv1303.5787B}), radio emission weeks to years after from interaction of the outflow with the interstellar medium~\cite{2011Natur.478...82N}, radio to X-rays on second to day timescales from shocks in binary NS mergers~\cite{Kyutoku:2012fv}, emission from tidally induced crust shattering~\cite{Tsang:2011ad,Tsang:2013mca}, and bursts from the eventual collapse of hypermassive NSs following binary NS mergers~\cite{Lehner:2011aa,Falcke:2013xpa}. It is thus easy to anticipate that being able to detect and identify properties of mergers will reveal a wealth of information: characterstics of source populations, testing general relativity in the dynamical strong field regime, and for mergers with NSs, learning about their composition and details of the electromagnetic emission mechanisms.

From a GW data analysis perspective, an important class of compact object coalescences are the so-called quasi-circular inspirals: binaries that are born with a sufficiently large periapse that GW emission will effectively circularize the orbit prior to it entering the LIGO band\footnote{For brevity we refer to the ground-based detector band as the ``LIGO'' band.}, regardless of the initial eccentricity of the orbit. Optimal template-based searches are well adapted to this class of binary, and significant progress has been made toward implementing them in current analysis pipelines~\cite{Aasi:2012rja}. There are also astrophysical mechanisms that can produce binaries that have high eccentricity while emitting in the LIGO band. These include dynamical capture via energy loss to GW emission during a close 2-body in a dense cluster~\cite{O'Leary:2008xt,lee2010short}, a merger induced during a binary-single star interaction in a similar environment~\cite{Samsing:2013kua}, and Kozai-resonant enhancement of eccentricity in a hierarchical triple system~\cite{Wen:2002km,Kushnir:2013hpa,Seto:2013wwa,Antognini:2013lpa,Antonini:2013tea}. Event rates are highly uncertain for both classes of binary (see~\cite{2010CQGra..27q3001A} for a review of estimates for quasi-circular inspiral rates, and~\cite{O'Leary:2008xt,lee2010short,Kocsis_Levin,Tsang:2013mca,east2013observing} for estimates and discussions of dynamical capture systems), though it is generally expected quasi-circular inspirals will be prevalent. Even if eccentric mergers are rare, they could be exquisite laboratories to learn about the physics of compact objects mergers mentioned above for several reasons. The GW emission is concentrated in bursts about periapse passage, where velocities are much higher than comparable emission frequencies in a quasi-circular inspiral. Hence more total power is emitted in the dynamical strong field regime of GR, which lacks constraints from existing observations and experiments. This further has the consequence that when NSs are involved there is more potential for matter disruption, ejection and formation of more massive accretion disks~\cite{bhns_astro_letter,bhns_astro_paper,Gold2011,Rosswog2012,nsns_astro_letter}. This can significantly affect electromagnetic emission processes, and, if these events are frequent enough, the heavy r-process element abundances in the universe~\cite{Rosswog2012,2013RSPTA.37120272R}.

For GW data analysis, these properties of high eccentricity mergers present several problems. Quasi-circular templates, unsurprisingly, do not perform well over much of parameter space in searches of simulated data, due to large mismatches with the eccentric waveforms~\cite{east2013observing,Huerta:2013qb}. Perturbative waveform generation methods have not yet been extended to sufficient accuracy to be used as templates for high eccentricity binaries (though see recent advances in this regard for the effective-one-body approach~\cite{Bini:2012ji}). And given that the majority of the energy emission occurs in concentrated bursts at high $v/c$, one can anticipate that these methods would need to be extended to quite high order in $v/c$ to obtain waveforms that have the requisite phase coherence for multiple burst mergers. With regard to full numerical solutions, the long timescales between bursts suggests it may be computationally impossible to generate accurate numerical template banks of entire inspiral-merger-ringdown (IMR) events for the Advanced-LIGO detector era.  

This brings us to the motivation behind the work presented here, to develop a {\em practical} search strategy that will increase the volume of the universe within reach of GW observatories for high eccentricity IMR events, as measured against existing quasi-circular inspiral and unmodeled burst searches. Our approach is to adapt the incoherent (power) stacking method introduced in~\cite{kalmus2009stacking} to search for GW emission coincident with soft gamma-ray repeater events. There, the observed time and duration of each gamma ray burst was used to define a sequence of time-frequency tiles in the GW data stream over which power would be integrated, and a statistically significant excess relative to detector noise searched for. Here, we use model IMR waveforms to inform the choice of timing for an ensemble of time-frequency tiles, and we search for excess power in the sum over each member of the ensemble. Thus we still need a high eccentricity ``template'' bank; however, the accuracy requirements are significantly less than what would be required for a matched filter search. Effectively, the model needs to predict the timing of each burst well enough to ensure that the majority of the power of the bursts occurs at those times to within a prescribed, tunable uncertainty interval. This requirement is much less stringent than the need to accumulate well less than a cycle in phase error, which is a requirement for efficient matched filtering. The downside of power stacking is it is obviously not as optimal as matched filtering. If each burst in a sequence is identical, power stacking would accumulate signal-to-noise ratio (SNR) as $N^{1/4}$ compared $N^{1/2}$ for matched filtering in an $N$-burst event. For IMR events the bursts do evolve with time, and in particular when folded in with model detector noise this scaling does not generally hold. Nevertheless, we show here that power stacking can still achieve a significant gain compared to single-burst searches.

The remainder of this work is structured as follows. In Sec.~\ref{sec:model}, we review the model, introduced in~\cite{east2013observing}, that we use to generate gravitational waveforms. The model follows equatorial geodesic motion on a Kerr BH background, coupled to quadrupole GW energy and angular momentum loss to evolve the parameters of the orbit with time. Though the model does not contain all the relevant parameters or likely the needed accuracy to be used for a ``real'' search, it captures enough of the physics of eccentric signals that it is adequate to use to develop the power stacking algorithm. In Sec.~\ref{sec:detection-prospects}, we describe some of the basics of  time-frequency search methods for unmodeled gravitational wave bursts that provide the building blocks for our targeted search. In Sec.~\ref{sec:qstack}, we introduce the power stacking method for searching for gravitational wave signals from high eccentricity IMR events. It is beyond the scope of this work to explore all the intricacies and details that would need to be worked out to develop this method into a mature search strategy. These include issues of parameter extraction and degeneracies, required accuracy of the waveform model and effects of modeling error, the efficacy of the method over the full range of relevant binary parameters, and analysis with multiple detectors. However, we do begin to answer some of these questions using Monte Carlo simulations over a limited range of parameters for a single detector with simulated noise. Our results suggest that power stacking could increase the detectable horizon distance by as much as a factor of three compared to single burst searches over much of the parameter space studied (but on the flip side, it is still a similar factor less than what could be achieved with matched filtering), and is quite robust to waveform modeling uncertainties. Finally, we discuss directions for future work in Sec.~\ref{sec:discuss}.

\section{Waveform model}\label{sec:model}

As mentioned in the introduction, neither existing full numerical nor perturbative methods of calculating highly eccentric IMR waveforms are likely to be accurate enough to be of use in an optimal matched filter search, motivating the search for alternative search strategies. Fortunately, to be able to develop new analysis methods and compare to matched filtering we do not need extremely accurate waveforms. Instead, the model waveforms need only capture the qualitative features of highly eccentric signals that are relevant to the data analysis. These waveforms then define the ``true'' signals  injected into simulated detector noise; likewise, they are used to define the optimal SNR that could be achieved with a matched filter search as the baseline to compare with the alternative approach. 

\begin{figure}
\includegraphics[width=.45\textwidth,draft=false]{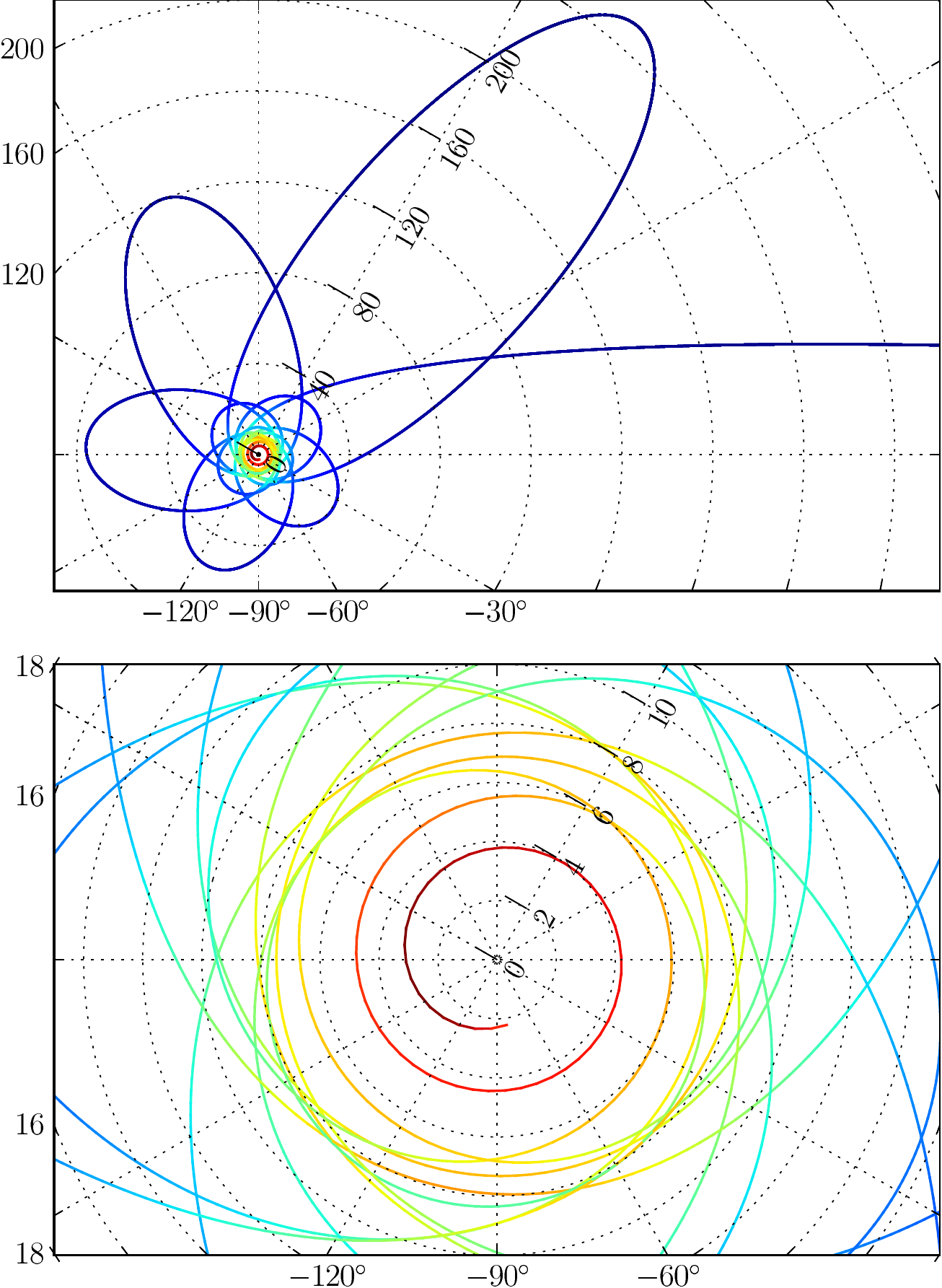}
%FP: replaced trajectory.pdf with the smaller file trajectory_sm.pdf. The latter's fonts
%don't look that great on screen on my system, but it print's OK, and shaves almost 1MB off the final pdf size.
%Might want to do the same with some of the other large figures (just did pdf2ps followed by ps2pdf)
\caption{Simulated trajectory of a binary with parameters $M=10\,M_{\astrosun}$, $r_{p}=8M$, mass ratio $q=1$, and initial eccentricity $e_{0} = 1$ as computed using the model described in the text. Colors correspond to the magnitude $\sqrt{h_{+}^{2} + h_{\times}^{2}}$ of emitted gravitational radiation, and the integration is stopped when $r$ reaches the innermost photon orbit. The radial coordinate is given in units of $M$. See Fig.~\ref{fig:waveformrp8} for the waveform corresponding to this orbit.}
\label{fig:trajectory-rp10}
\end{figure}

\begin{figure}
\includegraphics[width=.45\textwidth]{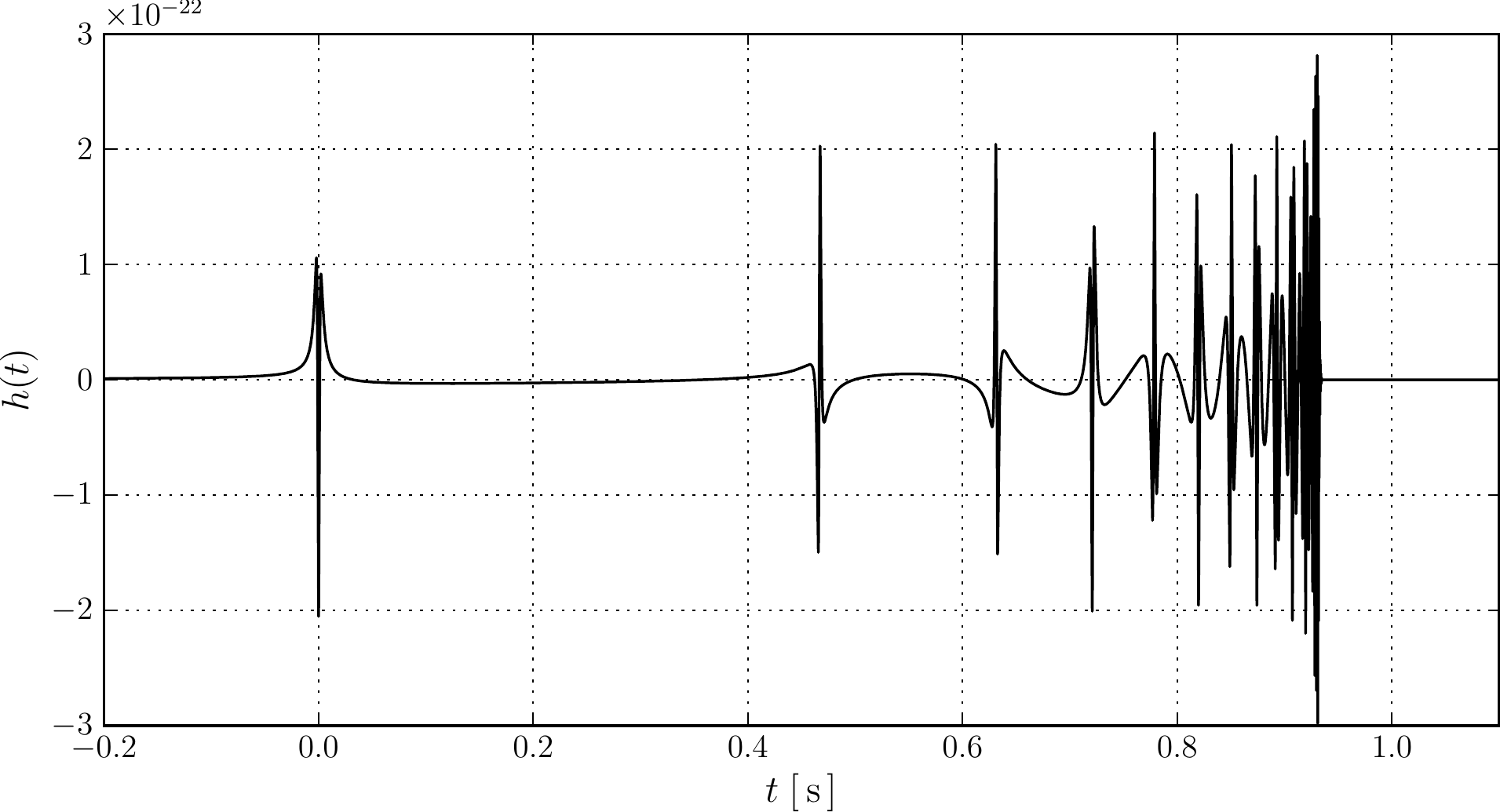}
\includegraphics[width=.45\textwidth]{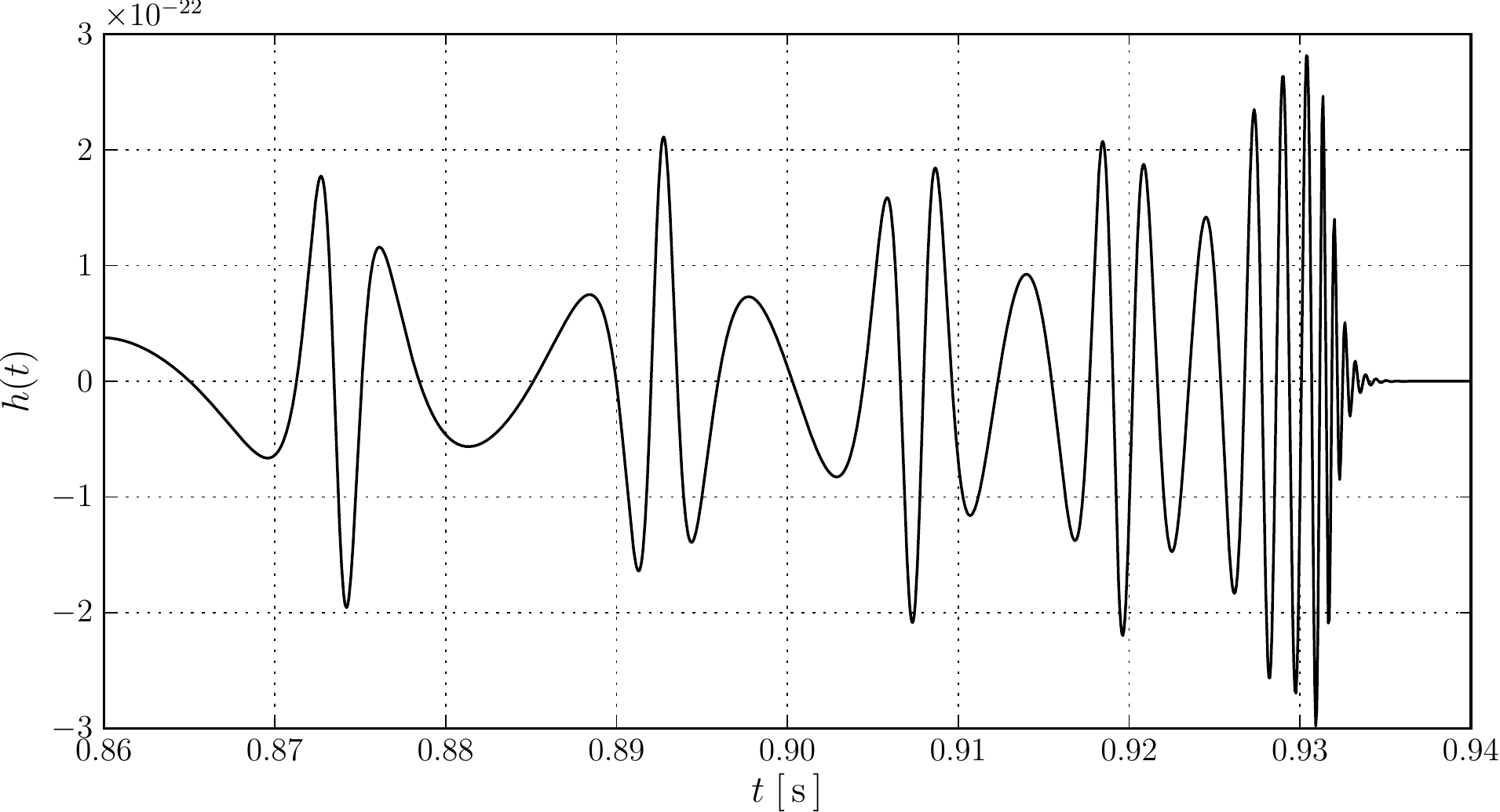}
\caption{Simulated gravitational wave strain $h(t)$ computed using the model described in the text for a binary with parameters $M=10\,M_{\astrosun}$, $r_{p}=8M$, $q=1$, $e_{0} = 1$ at $100\,\mathrm{Mpc}$ and optimally oriented. The bottom panel is a close-up about the time of merger of the signal shown in the top panel.}
\label{fig:waveformrp8}
\end{figure}

The binary IMR model we use was introduced before in~\cite{east2013observing}, though since it plays a central role in the rest of this paper we review it here (see Figs.~\ref{fig:trajectory-rp10} and \ref{fig:waveformrp8} below for a sample inspiral trajectory and corresponding gravitational waveform computed with this model). It is a hybrid composed of two components: an effective Kerr geodesic description of the inspiraling binary sourcing quadrupole radiation, and a merger-ringdown model originally developed for quasicircular systems. We describe these components separately below.

\subsection{Inspiral model}

We model the inspiral phase beginning with geodesic-{\em like} motion in an {\em effective} Kerr geometry. The total mass $M$ and angular momentum $J=a M$ of the Kerr geometry is set to the (instantaneous) net rest mass and sum of spin and orbital angular momenta of the binary respectively, while the parameters of the geodesic are the (instantaneous) reduced energy and orbital angular momentum of the binary. This is akin to an effective one body reduction, and the reason we include orbital angular momentum in the spin of the effective black hole is this was shown to better reproduce zoom-whirl like behavior in a similar geodesic model of equal mass merger simulation results~\cite{khurana}. Note though that in this study we only consider spinless binary components, so the spin of the effective Kerr black hole is entirely due to orbital angular momentum.

We solve the geodesic equation using the Boyer-Lindquist form of the Kerr metric:
\begin{eqnarray}
ds^{2} = &-&\left(1-\frac{2Mr}{\Sigma} \right) dt^{2} 
          + \frac{\Sigma}{\Delta} dr^{2} + \Sigma d\theta^{2} \nonumber \\
         &+& \left[r^2+a^2+ \frac{2Ma^2r\sin^2\theta}{\Sigma} \right] \sin^{2}\theta d\phi^{2}\nonumber\\
         &-&\frac{4Ma r\sin^{2}\theta}{\Sigma} dtd\phi,
\end{eqnarray} 
where:
\begin{align}
	\Sigma &= r^{2} + a^{2} \cos^{2}\theta , \nonumber\\
	\Delta &= r^{2} + a^{2} - 2Mr\nonumber.
\end{align} 
That the initial binary members have zero spin implies there will be no spin-induced orbital plane precession, hence the relevant geodesics are equatorial ($\theta=\pi/2$) and uniquely described by the two reduced energy and angular momentum parameters. We characterize each binary by the mass ratio $q$ ($q\in(0,1]$), the initial periapse $r_{p}$ and initial eccentricity $e_0$. At each time step of the geodesic integration we consider the black hole and geodesic parameters constant. To incorporate the consequences of gravitational wave emission we adjust these parameters between time steps as follows. First, we map the trajectory to the center of mass frame in a flat Cartesian space; specifically, if $m_{1}$ and $m_{2}$ are the masses of the two bodies in the binary, and $(r,\phi)$ are the Boyer-Lindquist coordinates of the geodesic, the Cartesian coordinates of the bodies are $\vec{x}_{1} = (m_{2}/M) \vec{R}$ and $\vec{x}_{2} = (m_{1}/M)\vec{R}$, with $\vec{R} = r(\cos\phi \; {\hat{x}} + \sin\phi \; {\hat{y}})$.  Then, using quadrupole formulas, we compute the waveform and energy and angular momentum radiated, using the latter two quantities to adjust the parameters of the geodesic and Kerr metric accordingly.  The integration is terminated when the radial coordinate $r$ reaches the innermost circular photon orbit (or ``light ring'') $r_{\mathrm{LR}}$,
\begin{equation}
	r_{\mathrm{LR}} = 2M \left[ 1 + \cos\left( \frac{2}{3} \cos^{-1}(-a) \right) \right],
\end{equation}
at which point we transition to the merger/ringdown waveform, discussed next.

The above inspiral model is clearly {\em ad hoc}, yet as described in~\cite{east2013observing} matches numerical results from single, small periapse ($r_p$) encounters quite well, and approaches the results from PN theory in the limit of large $r_p$.  

\subsection{Merger model}\label{sec:IRS}

The merger/ringdown phase is modeled using the \emph{implicit rotating source} (IRS) model \cite{baker2008mergers,kelly2011mergers}. The model approximates the gravitational radiation emitted by a coalescing binary using an effective rigidly rotating source with adiabatically-evolving multipole moments.  In the IRS model, the phase evolution of emitted gravitational waves asymptotically approaches the least-damped quasinormal mode frequency $\omega_{\mathrm{QNM}}$: 
\begin{align}
	\omega(t) &= \omega_{\mathrm{QNM}} (1 - \hat{f}), \\
	\hat{f} &= \frac{c}{2} \left( 1 + \frac{1}{\kappa} \right)^{1+\kappa} \left[ 1 - \left( 1 + \frac{1}{\kappa} e^{-2t/b} \right)^{-\kappa} \right],
\end{align} 
where $c$ and $\kappa$ are free parameters of the model, and $b = 2Q / \omega_{\mathrm{QNM}}$ is a function of the quality factor $Q$ of the final BH. The quality factor $Q$ is approximately the number of oscillations required for the energy of the oscillating system to be attenuated by a factor of $e^{2\pi}$. The parameters $c$ and $\kappa$ are fixed as constants that give good fits for a wide range of binary parameters.

The amplitude is modeled by: 
\begin{equation}
	A(t) = \frac{A_{0}}{\omega(t)} \left( \frac{\left| \dot{\hat{f}} \right|}{1 + \alpha(\hat{f}^{2} - \hat{f}^{4})} \right)^{1/2},
\end{equation} 
where $A_{0}$ is an overall amplitude factor and $\alpha$ is a free parameter that is chosen such that it is a good fit to numerical simulations (see \cite{east2013observing}).
 
\section{Time Frequency Analysis}
\label{sec:detection-prospects}

As illustrated in Fig~\ref{fig:waveformrp8}, the waveform of a typical high eccentricity binary begins in a repeated burst phase, with the instantaneous eccentricity steadily decreasing with each burst. If the initial periapse is sufficiently large, the orbit will eventually circularize. However, for much of parameter space relevant to dynamical capture~\cite{east2013observing}, binary-single interaction induced mergers~\cite{Samsing:2013kua} and Kozai-resonance driven mergers~\cite{Antognini:2013lpa,Antonini:2013tea} some fraction, if not all of the GW energy emitted within the LIGO band will come from the repeated burst phase. 

Matched filtering is the optimal detection strategy when the signal waveform can be precisely modeled. However, as discussed in the introduction, eccentric waveforms where most of the observable energy is within the repeated burst phase are poor candidates for matched filtering today. For full numerical simulations, this is because computational resources do not exist to produce accurate template banks for the full range of relevant parameters. Perturbative methods have not been developed to sufficiently high order to provide the requisite phase-accuracy over the lifetime of the typical signal.

On the other hand, time-frequency methods, though sub-optimal, can be effective for unmodeled or ``poorly'' modeled events. For GW searches to date these have only been developed for single, isolated bursts~\cite{Abadie:2010mt}, with the notable exception of a power stacking method targeting emission associated with soft gamma ray repeater events~\cite{kalmus2009stacking}. The goal of our work is to adapt this latter method to highly eccentric IMR events. In the remainder of this section we give an overview of time-frequency analysis, and specifically the Q-transform that underlies the power stacking procedure described in subsequent sections.

A time-frequency analysis begins by constructing a \emph{time-frequency representation} of the detector data $x(t)$. To do so,
we choose a two-parameter family of basis functions $\{ \psi(\tau, f) \}$ that covers a region of interest in time-frequency space. Then, the projection of $x(t)$ onto these basis functions is computed. For example, in the case of a short-time Fourier transform (STFT), the projections $X(\tau, f)$ are given by:
\begin{equation}
	X(\tau, f) = \int_{-\infty}^{\infty} x(t) w(t - \tau) e^{-2\pi i f t} \: dt,
\end{equation}
where $w(t)$ is a window function, such as a Hann window. The basis can be specialized if it is known that the class of target waveforms projects preferentially onto a particular family of functions.

Once a time-frequency representation has been constructed, bursts can be detected by searching for excess power in a time-frequency tile above the power expected for detector noise alone \cite{anderson2001excess}. If this excess power is above some threshold (set by the desired false alarm rate), an event is registered.  Examples of time-frequency methods include the TFCLUSTER algorithm \cite{sylvestre2002tfcluster}; Waveburst, which uses a wavelet decomposition \cite{klimenko2004wavelet}; and the Q-Pipeline \cite{chatterji2005search}.  We have chosen to use the Q-Pipeline as the basis for our search method for eccentric binaries, and so in the remainder of this section we provide some details on the Q-Pipeline.

\subsection{The Q-Transform}
\label{sec:qtransform}

\begin{figure}
\includegraphics[width=.45\textwidth]{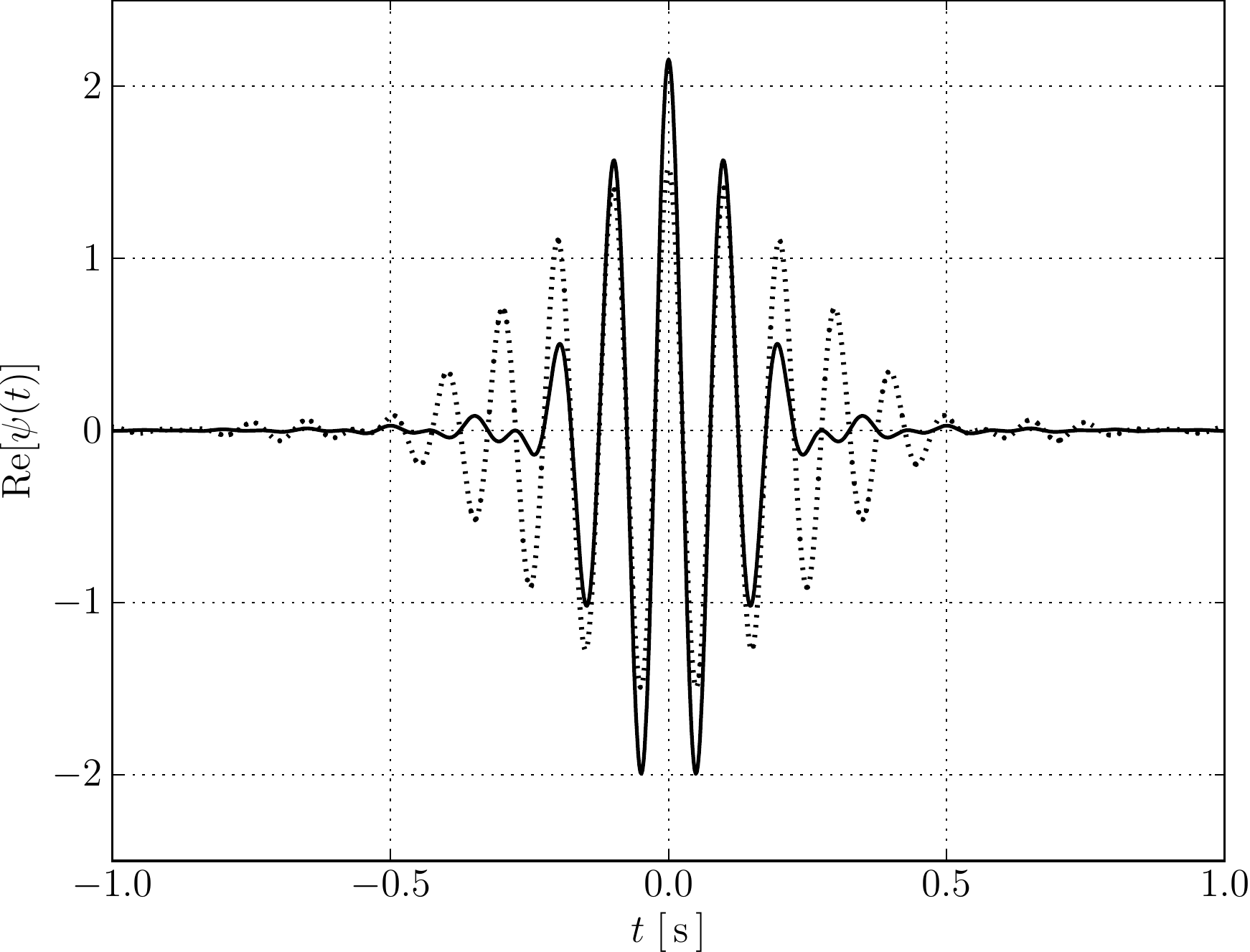}
\caption{The real part of two Q transform basis functions corresponding to parameters $\phi = 10\,\mathrm{Hz}$ and $\tau = 0\,\mathrm{s}$, with $Q = 10$ (solid line) and $Q = 20$ (dotted line). The functions have been normalized such that $\| \psi \| = 1$.}
\label{fig:qtransform-basis}
\end{figure}

The Q-Pipeline is one example of a time-frequency method used to search for poorly-modeled gravitational wave bursts. The key step in the method is the Q-transform, which is the projection of time series data onto a multiresolution basis of windowed complex exponentials.  Recall that we want to choose a family of basis functions that are each ``well-localized'' in time and frequency.  For any basis function $\psi(t)$, we define its characteristic center time $\tau$ and the characteristic center frequency $\phi$ as follows:
\begin{align}
	\tau &= \int_{-\infty}^{\infty} t|\psi(t)|^{2} \: dt, \\
	\phi &= 2 \int_{0}^{\infty} f |\tilde{\psi}(f)|^{2} \: df,
\end{align}
where in the definition of $\phi$ we include a factor of $2$ since we only integrate over positive frequencies. The squared uncertainty in time and frequency of $\psi(t)$ is defined by the corresponding variances:
\begin{align}
\label{eq:time-uncertainty}
	\sigma_{t}^{2} = \int_{-\infty}^{\infty} (t-\tau)^{2} |\psi(t)|^{2} \: dt, \\
\label{eq:freq-uncertainty}
	\sigma_{f}^{2} = 2 \int_{0}^{\infty} (f-\phi)^{2} |\tilde{\psi}(f)|^{2} \: df.
\end{align}
For bursts with no zero-frequency content, these quantities are constrained by an uncertainty relation \cite{gabor1946theory}:
\begin{equation}
	\sigma_{t} \sigma_{f} \, \geq \, \frac{1}{4\pi}.
\end{equation}
Moreover, this minimum uncertainty limit is achieved by Gaussian-windowed complex exponentials \cite{gabor1946theory}:
\begin{align}
\label{eq:minimum-uncertainty-function}
	&\psi(t) = \left( \frac{1}{2\pi \sigma_{t}^{2}} \right)^{1/4} \exp\left[ -\frac{(t-\tau)^{2}}{4\sigma_{t}^{2}} \right] \exp\Big[ 2\pi i \phi (t-\tau) \Big] \nonumber \\
		&= \left( \frac{8\pi \phi^{2}}{Q^{2}} \right)^{1/4} \exp\left[ -\frac{4\pi^{2}\phi^{2} (t-\tau)^{2}}{Q^{2}} \right] \exp\Big[ 2\pi i \phi (t-\tau) \Big],
\end{align}
where the dimensionless quality factor $Q$ (not to be confused with the quality factor used in the IRS ringdown model discussed in Sec.~\ref{sec:IRS}) is the ratio of center frequency to uncertainty in frequency:
\begin{equation}
	Q = \frac{\phi}{\sigma_{f}}.
\end{equation}
Intuitively, this $Q$ parameter can be understood as the number of oscillations of the windowed sinusoid, as illustrated in Fig.~\ref{fig:qtransform-basis}. Alternatively, this parameter can also be thought of as controlling the aspect ratio of $\psi(t)$ (as given by \eqref{eq:minimum-uncertainty-function}) in time-frequency space: ignoring the normalization factor, $\psi(t)$ in the low-Q limit is a delta function in time that is localized in time but not in frequency; in the high-Q limit, we recover the familiar Fourier basis that is localized in frequency but not in time.

\subsection{The Q-transform basis}
\label{sec:qtransform-basis}

In a general search for unmodeled gravitational wave bursts, a basis of minimum uncertainty functions described by \eqref{eq:minimum-uncertainty-function} allows the time-frequency structure of an arbitrary burst to be maximally resolved. Since one then does not have \emph{a priori} knowledge of the duration of a burst, the Q-transform uses a \emph{multiresolution basis} that resolves structure over multiple characteristic time and frequency scales. For the basis of windowed complex exponentials, this is controlled by the $Q$ parameter. Therefore, the Q-transform is a projection onto a three-parameter family of basis functions $\{ \psi(t;\tau, \phi, Q) \}$.

Note that this multiresolution basis is in general not orthogonal. However, since the goal of the Q-pipeline is burst detection rather than signal reconstruction, the lack of orthogonality is not a significant concern. The only caveat is that we may need to be careful in accounting for statistical correlation between projections onto overlapping basis functions.

In choosing the number and placement of basis functions to cover the targeted parameter space, we want any well-localized burst in the targeted parameter space to be a good match to some basis function in order to maximize the chances of detection. Simultaneously, the chosen basis should be as sparse as possible in the interest of computational efficiency.  Let $\Psi$ be the space of normalized Gaussian-windowed complex exponentials within some target parameter space $[\tau_{\min}, \tau_{\max}] \times [\phi_{\min}, \phi_{\max}] \times [Q_{\min}, Q_{\max}]$. The tradeoff between the competing goals described above is parameterized by the mismatch parameter $\mu_{\mathrm{max}}$, which we define for a given (finite) basis $\{ \psi \} \subset \Psi$ as:
\begin{equation}
	\mu_{\mathrm{\max}}(\{ \psi \}) = \max_{\varphi \in \Psi} \min_{\psi \in \{ \psi \}} \Big( 1 - \big|\langle \varphi \, | \, \psi \rangle \big|^{2} \Big),
\end{equation} 
i.e., the worst case energy loss due to mismatch between a well-localized burst $\varphi \in \Psi$ and the best-match basis function $\psi \in \{ \psi \}$. Intuitively, a sparser basis leads to a greater value of $\mu_{\max}$ and vice versa.

To select a basis, we use the following procedure: first, we fix some value $\mu > 0$, and construct a basis $\{ \psi \}$ such that $\mu_{\mathrm{\max}}(\{ \psi \})\, \leq\, \mu$ over the target parameter space. It can be shown that this bound is achieved by a basis that is logarithmically distributed in $Q$, logarithmically distributed in $\phi$, and linearly distributed in $\tau$ \cite[Section~3.2.2]{chatterji2005search}.

\subsection{Statistics}
\label{sec:qtransform-statistics}

We assume that the detector noise $n(t)$ is stationary white noise that is Gaussian-distributed. The normalized energy $Z$ for a basis function $\psi$ is then defined as
\begin{equation}
	Z = \frac{| \langle \psi \,|\, x \rangle |^{2}}{\mathrm{E}\big[ \big| \langle \psi \,|\, n \rangle \big|^{2} \big]},
\end{equation}
where $x$ is the detector output and $E[\cdot]$ is the expectation value of the argument. When the Q-transform is applied to stationary white noise, this normalized energy is exponentially distributed with unit mean \cite{chatterji2005search}:
\begin{equation}
	f(Z)\, dZ = \exp(-Z)\,dZ.
\end{equation}
This normalized energy can be related to the SNR obtained via matched filtering. Let $x(t) = s(t) + n(t)$, where $s(t)$ denotes the signal. Let $X = \langle \psi \, | \, x \rangle $, $S = \langle \psi \, | \, s \rangle$ and $N = \langle \psi \, | \, n \rangle$.  Then $| X |^{2} = | S |^{2} + | N |^{2} + |S| |N| \cos \theta$, where $\theta$ is a uniformly distributed phase term. Therefore $\mathrm{E}[|X|^{2}] = |S|^{2} + \mathrm{E}[|N|^{2}]$, and
\begin{equation}
	\mathrm{E}[Z] = \frac{\mathrm{E}[|X|^{2}]}{\mathrm{E}[|N|^{2}]} = \frac{|S|^{2}}{\mathrm{E}[|N|^{2}]} + 1.
\end{equation} 
If the (two-sided) power spectral density $S_{n}(f)$ varies slowly over the bandwidth of $\psi$, then $\mathrm{E}[|N|^{2}] \approx 2S_{n}(\phi)$, where $\phi$ is the center frequency of $\psi$. We recognize $|S|^{2}/S_{n}(\phi)$ as the square of the SNR obtained via matched filtering using the template $\psi$ (again, to within the approximation that $S_{n}$ varies slowly). Therefore, if $s(t)$ is a sine-gaussian burst with optimal SNR $\rho$, the normalized energies $Z$ satisfy the relation:
\begin{equation}
\label{eq:normalized-energy-snr}
	\mathrm{E}[Z] = \frac{1}{2} \rho^{2} + 1.
\end{equation}
The factor of $1/2$ can be interpreted as the ``cost'' of projecting over complex basis functions, which is done since the phase of the signal is unknown.

\subsection{Example}

\begin{figure}
\includegraphics[width=.45\textwidth]{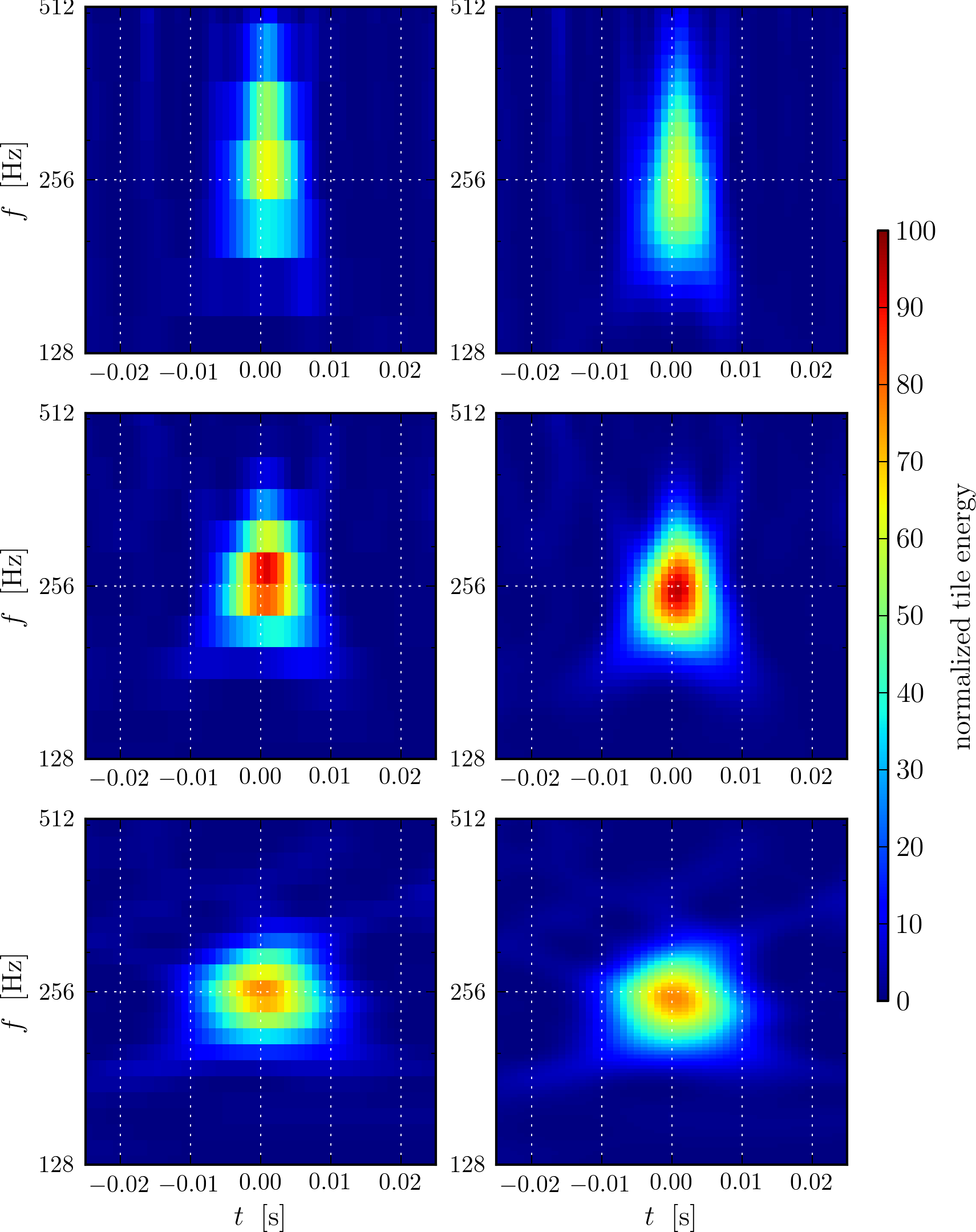}
\caption{An example of the Q-transform applied to a sine-gaussian burst ($\tau=0$, $\phi=256\,\mathrm{Hz}$, $Q=8$) injected in LIGO noise at $\rho = 14$. Top: $Q=4$. Middle: $Q=8$. Bottom: $Q=16$. Left: $\mu=0.2$. Right: $\mu=0.01$.}
\label{fig:qtransform-example}
\end{figure}

As an example application of the Q-transform, we inject a sine-gaussian burst $h(t) = h_{0} \exp\left[ -4\pi^{2}\phi^{2} (t-\tau)^{2}/Q^{2} \right] \exp\left[ 2\pi i \phi (t-\tau) \right]$ with $\tau = 0$, $\phi=256\,\mathrm{Hz}$, and $Q=8$ into simulated noise characteristic of Initial LIGO. The burst is injected with optimal SNR $\rho = 14$. The resulting time series is Q-transformed with mismatch parameters $\mu = 0.2$ and $\mu = 0.01$. For ease of comparison between the two values of $\mu$, we choose fixed Q-planes of $Q=4,8,16$, though in practice these values will depend on $\mu$. The result is shown in Fig.~\ref{fig:qtransform-example}.

The lower value of $\mu$ results in a higher-resolution time-frequency image at the cost of computation time. Note that the highest observed $Z$ values (in the $Q=8$ transforms) are $Z \approx 100$, as expected from the SNR of the injected signal and \eqref{eq:normalized-energy-snr}.

\section{Detecting gravitational waves from highly eccentric binaries}
\label{sec:qstack}

Since existing methods of gravitational wave detection are poorly-suited for detecting signals emitted by highly eccentric binaries, we seek a detection algorithm that:
\begin{enumerate}
	\item Improves on existing search methods for unmodeled, well-localized bursts.
	\item Is more robust than matched filtering to mismatches between the template and the signal.
\end{enumerate}

In this section, we describe a novel search method that specifically targets gravitational waves from eccentric binaries. We first give a general outline of the method before describing an implementation based on the Q-transform, and then discuss the statistical properties of the method and give an example application. Finally, we characterize the properties and effectiveness of the method using Monte Carlo simulations.

The main idea is a variant of excess power methods, though instead of identifying a single time-frequency tile whose power exceeds a certain threshold (as is done in searches for well-localized bursts), we instead sum excess power over a set of tiles. In particular, we do not require that the signal be detectable in any single time-frequency tile. This ``power stacking'' strategy has been used to search for gravitational wave signals from soft gamma repeater bursts by aligning repeated bursts with electromagnetic triggers \cite{abbott2009stacked,kalmus2009stacking}. In our method, tile selection is guided by our waveform model described above. A candidate event is identified when the computed statistic exceeds a predetermined threshold chosen to achieve a given false alarm rate.

Note then that our method still requires a bank of ``templates'' that informs which sequence of tiles is viable for the class of binaries being searched for. In other words, 
even though an individual sequence could
have as few as one (``direct collision'') to up to thousands of distinct tiles depending on the initial periapse, the number of tile-sequences that are searched over for any given starting time in the data stream is determined by the fundamental (intrinsic) parameters describing the binary (here, $M, r_p, e_0$ and $q$) and the extrinsic parameters describing the relative orientation of the binary and detector. For the analysis below it will be useful to have actual waveforms, and not just the corresponding set of times for the time-frequency tiles, 
though if implemented in a search pipeline only the latter would be needed.  Furthermore, we allow a tolerance for timing error due to the intrinsic uncertainty of
time-frequency analyses, but we could also vary this tolerance to account for an imperfect model; in this way, we could in principle use a
cruder strategy for choosing the sequence of times.

\subsection{Description}
\label{sec:qstack-description}

Here, we give a general framework for our approach that admits many possible implementations. Let $\{ \psi \}$ be a complete, but not necessarily orthogonal, basis for the Hilbert space of square-integrable functions on the real line, $L^{2}(\mathbb{R})$, where the basis functions $\psi$ are themselves absolutely- and square-integrable.  This last condition ensures that the basis functions are ``localized'' in time (as opposed to, for example, the familiar Fourier basis of complex exponentials, which are only localized in frequency). On this space, we define the time-domain inner product and $L^{2}$ norm:
\begin{equation}
	\langle f | g \rangle = \int_{-\infty}^{\infty} f(t) g^{*}(t) \: dt \quad \text{and} \quad \| f \| = \sqrt{\langle f | f \rangle}.
\end{equation}
For simplicity, let the basis functions be normalized such that $\| \psi \|  =  1$. Possible choices of basis functions include Gaussian-windowed complex exponentials and a wide selection of existing orthonormal wavelet families (for future work it would be interesting to investigate wavelets adapted to the waveform structure of eccentric binary bursts).

We assume we are given a model that provides waveforms $h(t)\in L^{2}(\mathbb{R})$ (to some level of accuracy) for the binary parameters of interest, here $(M, r_{p}, q, e_0)$. Let $h'(t)$ denote the ``whitened'' waveform:
\begin{equation}
	h'(t) = \int_{-\infty}^{\infty} \frac{\tilde{h}(f)}{\sqrt{S_{n}(f)}} \: e^{2\pi i f t} \: df,
\end{equation} 
where again $S_{n}(f) = S_{n}(-f)$ is the two-sided power spectral density of the detector noise $n(t)$, which we assume is stationary, and $\tilde{h}(f)$ is the Fourier transform of $h(t)$. The waveform is said to be ``whitened'' since the application of the whitening kernel $S_{n}(f)^{-1/2}$ to stationary detector noise $n(t)$ reduces its power spectral density to a constant function of frequency.
	
With respect to the choice of basis, we define the \emph{$N$-burst signature} (or for brevity {\em signature}) of a set of eccentric binary parameters, $\mathcal{S}_{N}(M, r_{p}, q, e_0)$ to be the set of $N$ distinct basis functions $\{ \psi_{1}, \dots, \psi_{N} \}$, such that the sum
\begin{align}
	\mathcal{C} &= \sum_{i=1}^{N} \frac{|\langle \psi_{i} | h' \rangle |^{2}}{\mathrm{E}\big[|\langle \psi_{i} | n' \rangle|^{2}\big]} \\
		&= \sum_{i=1}^{N} \big|\langle \psi_{i} | h' \rangle\big|^{2}
\end{align}
is maximized. The sum is weighted by the expected value of noise power projected onto the the basis function $\psi_{i}$, but since $n'(t)$ is whitened noise and $\| \psi_{i} \| = 1$, we have that $\mathrm{E}\big[|\langle \psi_{i} | n' \rangle|^{2}\big] = 1$. The signature $\mathcal{S}_{N}$ is therefore the set of $N$ basis functions that best approximates the whitened signal $h'(t)$. The number of basis functions $N$ can either be a fixed parameter or be determined by the signature generation algorithm.

Suppose that the model waveform $h(t)$ for a given, single set of parameters captures general features of \emph{actual} signals over some region of parameter space, even if $h(t)$ does not exactly match the actual signal produced by a binary with the same initial parameters. Let $D:\Psi \times \Psi \rightarrow [0, +\infty)$ be a function that satisfies the usual conditions for a metric on the space of basis functions $\Psi$. For example, a natural choice for $D$ is the Euclidean distance $D(\psi_{1}, \psi_{2}) = \| \psi_{1} - \psi_{2} \|$.  For each $i \in \{1, \dots, N \}$, let $\xi_{i} \, \geq \, 0$ be a real parameter that defines the ``slack'' for each best-match basis function $\psi_{i}$. More precisely, let $B_{i} = \{ \varphi : D(\varphi, \psi_{i}) \,\leq \, \xi_{i} \}$ be the set of basis functions within a distance $\xi_{i}$ of $\psi_{i}$ as measured by $D$. 

To apply the signature at a time offset $\tau$ to whitened detector data $x'(t)$, we compute the sum:
\begin{align}
	\mathcal{E}(\tau) &= \sum_{i=1}^{N}  \max_{\varphi_{i} \in B_{i}} \Big|\langle \varphi_{i}(t) | x'(t + \tau)  \rangle \Big|^{2} \nonumber \\
		&=  \sum_{i=1}^{N} \max_{\varphi_{i}\in B_{i}} \left| \int_{-\infty}^{\infty} \varphi_{i}(t) x'^{*}(t + \tau) \: dt \right|^{2}.
\end{align}

The maximization of the projection of $x'(t)$ onto basis functions in a neighborhood of each $\psi_{i}$ captures the intuition that we want to be able to detect waveforms that are ``near'' the simulated waveform. As with other time-frequency searches, we register an event at some time $\tau$ if the statistic $\mathcal{E}(\tau)$ exceeds a predetermined threshold value that is chosen to achieve a desired upper limit on the false alarm. We discuss event thresholds in Sec.~\ref{sec:qstack-significance}.

There is a tradeoff between the $\mathcal{E}$ threshold necessary to achieve a given false alarm rate and the robustness of the search to parameter mismatch and modeling error. We discuss the statistics of this search method in Sec.~\ref{sec:qstack-statistics}. In the following section, we describe a straightforward implementation of this method.

\subsection{Implementation}
\label{sec:qstack-implementation}

We implement the method described in Sec.~\ref{sec:qstack-description} as an extension of the Q-transform. This choice is motivated by the following considerations: (i) by performing a one-time Q-transform, we can easily apply multiple signatures to a time series, (ii) the overcomplete, multiresolution Q-transform basis increases our prospects of finding good matches to the signature, and (iii) the Q-transform is an established method that has been deployed in past LIGO science runs. 

Recall that the Q-transform basis of Gaussian-windowed complex exponentials minimizes the product of $\sigma_{t}$ and $\sigma_{f}$ (as defined in \eqref{eq:time-uncertainty} and \eqref{eq:freq-uncertainty}), such that $\sigma_{t}\sigma_{f} = 1/4\pi$. In the following, we define the characteristic duration $\Delta t$ and characteristic bandwidth $\Delta f$ of a basis function as follows:
\begin{align}
	\Delta t &= 2 \pi^{1/2} \sigma_{t} \\
	\Delta f &= 2 \pi^{1/2} \sigma_{f},
\end{align}
such that the product $\Delta t \Delta f$ is unity.  We now consider the signature generation and the signal analysis steps separately.

\subsubsection{Signature generation}
\label{sec:signature-generation}

\begin{figure}
\includegraphics[width=.45\textwidth]{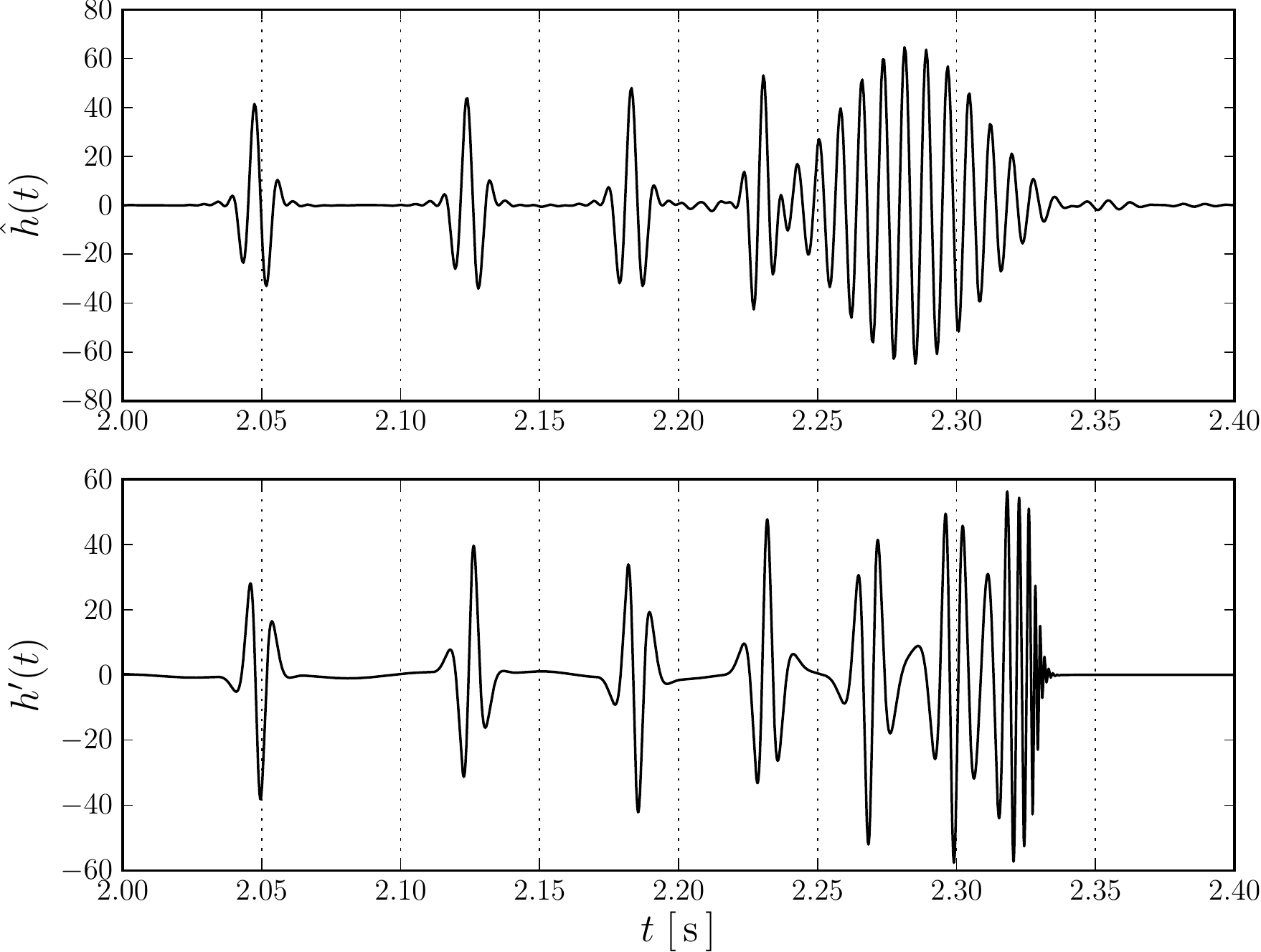}
\caption{Part of a whitened waveform $h'(t)$ (bottom panel) and the best-match approximation (top panel) as computed using the algorithm from Sec.~\ref{sec:signature-generation}. Note that the final burst in the approximation is a superposition of two basis waveforms. The binary parameters in this example are $M=25$, $r_{p} = 8$, $q=1$, and $e_0 = 1$.}
\label{fig:qstack-approx}
\end{figure}

Given binary parameters $(M, r_{p}, q, e_0)$, Q-transform parameters $(Q_{\min}, Q_{\max}, \phi_{\min}, \phi_{\max})$, and mismatch parameter $\mu$ (see Sec.~\ref{sec:qtransform-basis}), we compute a signature using the following procedure:
\begin{enumerate}
\item Using the waveform model, simulate (a) the emitted signal $h(t)$ for the given binary parameters and (b) the orbital separation $r(t)$ from $t=0$ until $t=t_{\mathrm{merge}}$, defined such that $r(t_{\mathrm{merge}}) = r_{\mathrm{LR}}$, the radius of the light ring. Assume for simplicity that the $+$ and $\times$ polarizations contribute equally, so that $h(t) = h_{+}(t) + h_{\times}(t)$ (for purposes of signature generation, the overall scaling of the signal is irrelevant).
\item Identify the sequence of times $t_{1}, t_{2}, \dots, t_{N}$ where $r(t)$ is at a local minimum. Let $t_{N} = t_{\mathrm{merge}}$. Therefore, times $t_{1}, \dots, t_{N-1}$ correspond to periapse passages of the binary, and $t_{N}$ corresponds to the merger. Let the number of basis functions $N$ equal the number of minima.
\item Compute $h'(t)$, the whitened signal.
\item Q-transform $h'(t)$ to give a set of time-frequency planes of fixed $Q$.
\item For $k=1,\dots,N$: \\
Find the basis function $\psi_{k}$ that maximizes $\big| \langle \psi_{k} \, | \, h' \rangle \big|^{2}$, subject to the constraint that $t_{k} \in [\tau_{k} - \Delta t_{k} / 2, \tau_{k} + \Delta t_{k} / 2]$, where $\tau_{k}$ is the center time of $\psi_{k}$ and $\Delta t_{k}$ is the characteristic duration of $\psi_{k}$.
\item Shift center times such that the time of the first basis function is $\tau_{1} = 0$. Return the set $\mathcal{S}_{N} = \{ \psi_{1}, \dots, \psi_{N} \}$.
\end{enumerate}

An example of a computed signature is shown in Fig.~\ref{fig:qstack-approx}. Note that in the figure, the real and imaginary components of the projection are expressed as an overall magnitude and phase; for the purposes of computing a signature, we are not interested in the real and imaginary projections, only in the energy $\big| \langle \psi_{k} \, | \, h' \rangle \big|^{2}$.

\begin{figure}
\includegraphics[width=.45\textwidth]{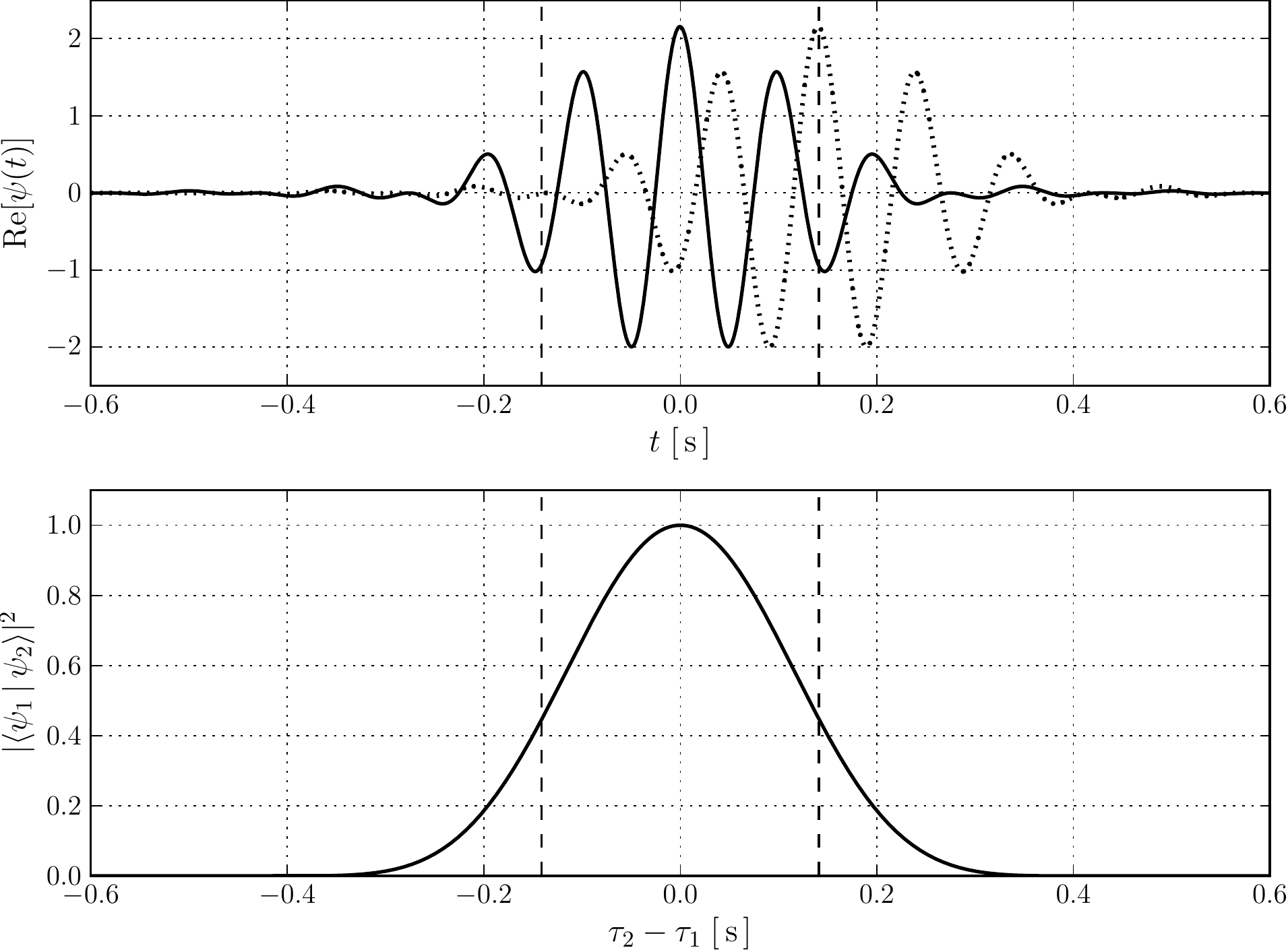}
\caption{Overlap between two basis functions $\psi_{1}$, $\psi_{2}$ with center times $\tau_{1}$, $\tau_{2}$ offset by $\Delta t = 2\pi^{1/2} \sigma_{t}$. In this case, $\big|\langle \psi_{1} \, | \, \psi_{2} \rangle\big|^{2} \approx 0.45$.}
\label{fig:qstack-overlap}
\end{figure}

\subsubsection{Signal analysis}
\label{sec:signal-analysis}

In the following, we let $Q_{k}$, $\tau_{k}$, $\phi_{k}$, $\Delta t_{k}$ and $\Delta f_{k}$ denote the respective quantities associated with $\psi_{k}\in\mathcal{S}_{N}$.  For a given detector output $x(t)$ of finite duration, a signature $\mathcal{S}_{N}$, Q-transform parameters $Q_{\min}, Q_{\max}, \phi_{\min}, \phi_{\max}$, a value for $\mu$, and a step size $\delta t$, we compute the $\mathcal{E}$ statistic at each time step using the following procedure:

\begin{enumerate}
\item Compute $x'(t)$, the whitened detector output.
\item Q-transform $x'(t)$.
\item For $T = j\cdot\delta t$, $j=0,1,2,\dots$:
\begin{enumerate}
	\item For $k=1,\dots,N$: \\
	Choose the time-frequency plane in the Q-transform of $x'(t)$ with Q value closest to $Q_{k}$. Let $c_{k}$ be the maximum energy of a time frequency tile that lies within a distance of $\Delta t_{k} / 2$ in time or $\Delta f_{k} / 2$ in frequency: $c_{k} = \max_{\psi} \big| \langle \psi \, | \, x' \rangle \big|^{2}$ for basis functions $\psi$ in this Q-plane, subject to the constraint that $\tau\in[\tau_{k} + T - \Delta t_{k}/2, \tau_{k} + T + \Delta t_{k}/2]$ and $\phi \in [\phi_{k} - \Delta f_{k}/2, \phi_{k} + \Delta f_{k}/2]$.
	\item Set $\mathcal{E}(T) = \sum_{k=1}^{N} c_{k}$.
\end{enumerate}
\end{enumerate}

We note that our power stacking method can be implemented efficiently. If we let $n$ be the number of samples in $x(t)$, the Q-transform is precomputed in $O(n\log n)$ time \cite{chatterji2005search}. At each time step, the computation of the statistic $\mathcal{E}(T)$ takes time $O(NM)$, where $N$ is the number of time-frequency tiles in the signature and $M=M(\mu)$ is an upper bound on the number of overlapping tiles. The set of overlapping tiles can be found efficiently if the Q-planes are represented as jagged arrays with rows of constant frequency. The number of overlapping tiles scales with the density of basis functions, and therefore increases as the mismatch parameter $\mu$ decreases. For $\mu = 0.2$, we observe $M\approx 40$.  The algorithm described above can easily be generalized to the case where we wish to search over multiple signatures $\{ \mathcal{S}^{(1)}, \dots, \mathcal{S}^{(m)} \}$ for a one-time precomputation of the Q-transform.  In our implementation, we consider tiles that fall within the time-frequency rectangle defined by the duration $\Delta t_{k}$ and bandwidth $\Delta f_{k}$ of $\psi_{k}$, so the metric and the slack parameters $\xi_{k}$ are chosen implicitly. Additionally, for each $\psi_{k}$ we consider only tiles within a single Q-plane.

\subsection{Statistics}
\label{sec:qstack-statistics}

\subsubsection{Distribution of $\mathcal{E}$ statistic}

We now discuss the statistical distribution of the $\mathcal{E}$ statistic when a signature is applied to detector noise in the absence of a signal. We will use this distribution to determine the significance of a given value of $\mathcal{E}$.

We again consider stationary white noise with zero mean and unit standard deviation. As discussed in Sec.~\ref{sec:qtransform-statistics}, the normalized energies $Z$ for Q-transform coefficients are exponentially distributed with unit mean: \begin{equation}
	f(Z) = e^{-Z}.
\end{equation} Given $N$ independent tiles, the distribution of the sum of their respective energies $Y = \sum_{i=1}^{N} Z_{i}$ is given by the Erlang distribution: \begin{equation}
	f(Y) = \frac{Y^{N-1} e^{-Y}}{(N-1)!}.
\end{equation} 
This would be the distribution of the $\mathcal{E}$ statistic if the maximization step in the algorithm is not performed, for example if we simply chose the nearest neighbor tile for each $\psi_{k}$ subject to the constraint that the chosen basis functions are orthogonal.

Instead, for each $\psi_{k}$, we choose the maximum energy over the set of overlapping tiles. The $\mathcal{E}$ statistic is distributed as the sum of $N$ maxima, each taken over at most $M$ exponentially distributed random variables that are not necessarily independent. There is no simple analytic expression for this distribution, so we approximate it for a given signature by fitting a shifted Gamma distribution to an ensemble of $\mathcal{E}$ statistics obtained by applying the algorithm to Gaussian white noise. The Gamma distribution is a two-parameter family of continuous probability distributions, with probability density function $f$ described by:
\begin{equation}
	f(x;\alpha, \beta) = \frac{\beta^{\alpha}x^{\alpha-1}e^{-\beta x}}{\Gamma(\alpha)}.
\end{equation}
We include a shift parameter $\mathcal{E}_{o}$ by letting $x = \mathcal{E} - \mathcal{E}_{o}$. The parameters $\alpha$, $\beta$ and $\mathcal{E}_{o}$ are then computed using maximum likelihood estimation with respect to an ensemble of $\mathcal{E}$ statistics over noise.

Even if projections onto distinct basis functions were independent, we cannot determine a closed-form expression for the distribution of $\mathcal{E}$. Suppose an orthonormal basis were chosen instead of the Q-transform basis, and assume for simplicity that $|B_{k} | = M$ for all $k$ and that the squared magnitude of the projection $Z$ is distributed exponentially as with the Q-transform. In this case, orthonormality guarantees that the projections onto the basis functions are independent and identically distributed. For any $B_{k}$, the cumulative distribution function $F$ of the maximum $Z_{\mathrm{max}}$ over the $M$ projections is given by: \begin{align}
	F(Z') &= \mathrm{Pr}[Z_{\mathrm{max}} \,\leq \, Z'] \nonumber \\
	&= \mathrm{Pr}[Z \, \leq \, Z']^{M}  \nonumber \\
	&= (1 - e^{-Z'})^{M}.
\end{align} The probability density function is then: \begin{equation}
	f(Z_{\mathrm{max}}) = M e^{-Z_{\mathrm{max}}} \left(1 - e^{-Z_{\mathrm{max}}} \right)^{M-1}.
\end{equation}
The distribution of the sum of the $N$ maxima is given by the $N$-fold convolution of the above distribution, which does not have a simple closed-form expression.

\subsubsection{Significance and detection thresholds}
\label{sec:qstack-significance}

From the best-fit distribution $\mathcal{D}$, we can determine the statistical significance of a given value of $\mathcal{E}$. The significance threshold for registering a detection is chosen as a function of the expected number of false events per year $N_{\mathrm{false}}$. For a step size of $\delta t$ and a significance threshold $\eta$, the expected number of false events in an observation period $T$ is:
\begin{equation}
	N_{\mathrm{false}} = \eta \frac{T}{\delta t}.
\end{equation} 
For example, a false alarm rate of $1\,\mathrm{yr}^{-1}$ and a step size of $10\,\mathrm{ms}$ requires a threshold of $\eta \approx 3.2 \times 10^{-10}$. It is often more convenient to express this as a negative log significance: $-\ln\eta \approx 22$. Using the cumulative distribution function of $\mathcal{D}$ and this threshold significance, we can compute the threshold statistic $\mathcal{E}^{*}$ for registering a detection.

It is worth emphasizing that these thresholds and false alarm rates depend on the assumption of stationary Gaussian noise. In practice, the false alarm rate is dominated by various non-stationary and non-Gaussian sources of noise, resulting in observed false alarm rates orders of magnitude higher than theoretically predicted \cite{abbott2007tuning}. However, false detection rates based on stationary Gaussian noise are a useful common standard for comparing the performance of different methods. We use fixed values of the false detection rate for our comparisons in Sec.~\ref{sec:qstack-simulation}. 

\subsection{Example}
\label{sec:qstack-example}

\begin{figure}
\includegraphics[width=.45\textwidth]{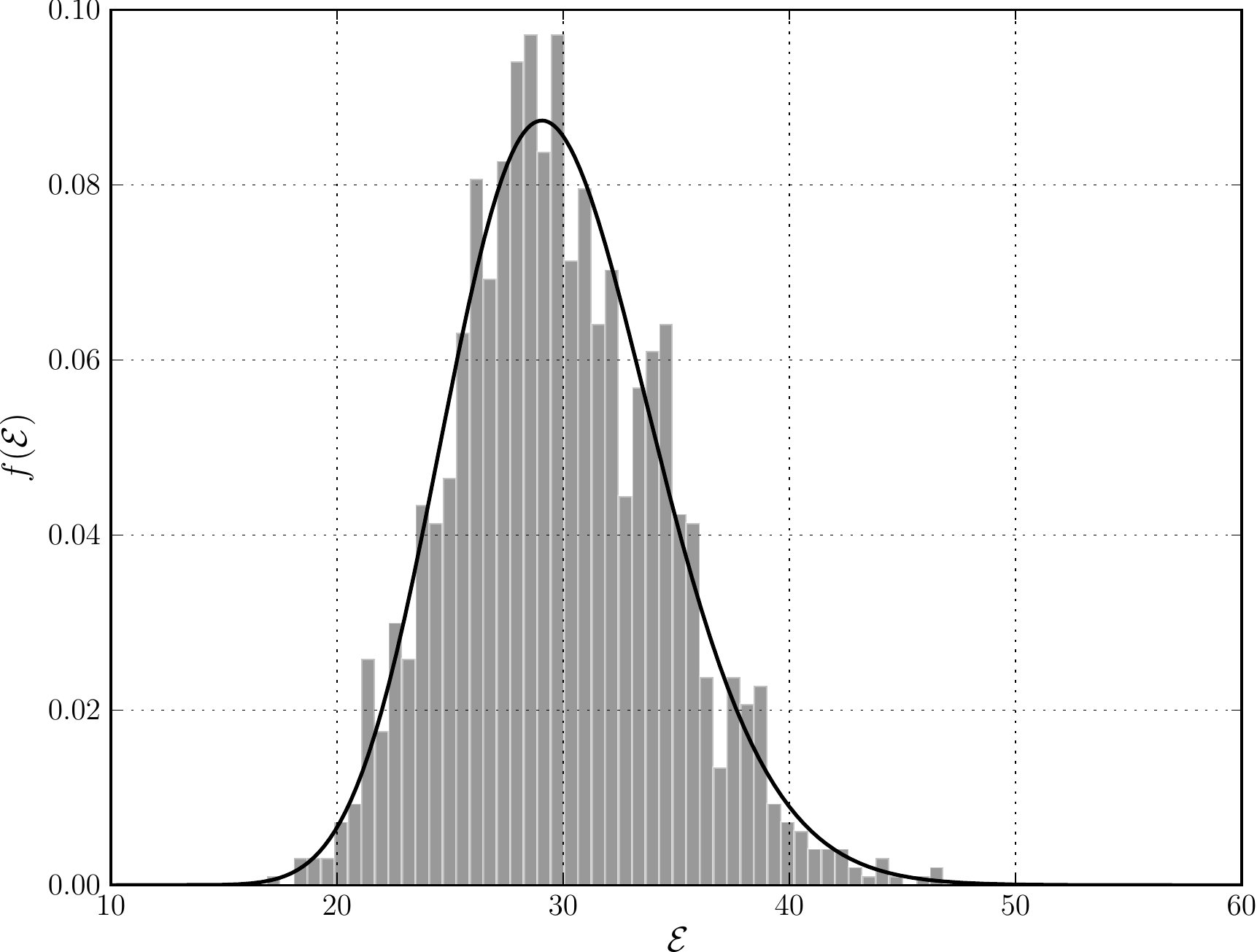}
\caption{Gamma distribution fit to an ensemble of $\mathcal{E}$ statistics on Gaussian white noise. The binary parameters used for computing the signature are $M = 10\,M_{\astrosun}$, $r_{p} = 8M$, and $q = 1$.}
\label{fig:qstack-distribution}
\end{figure}

\begin{figure}
\includegraphics[width=.45\textwidth]{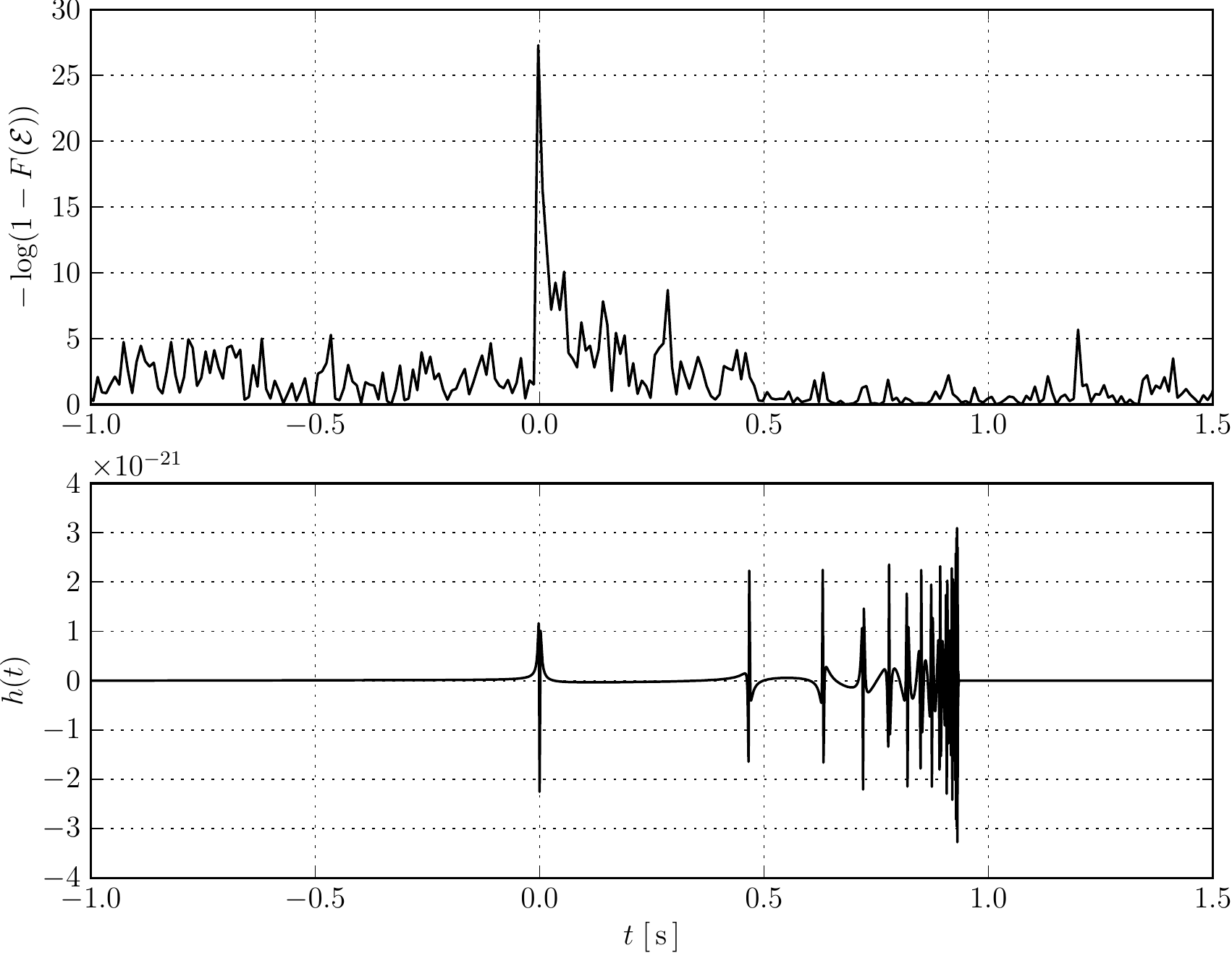}
\caption{The power stacking method in the matched case for the signal shown in the bottom panel, and parameters as decribed in the caption of Fig.~\ref{fig:qstack-distribution}. The top panel shows the statistical significance of the computed statistic $\mathcal{E}$ plotted against time.}
\label{fig:qstack-example-match}
\end{figure}

\begin{figure}
\includegraphics[width=.45\textwidth]{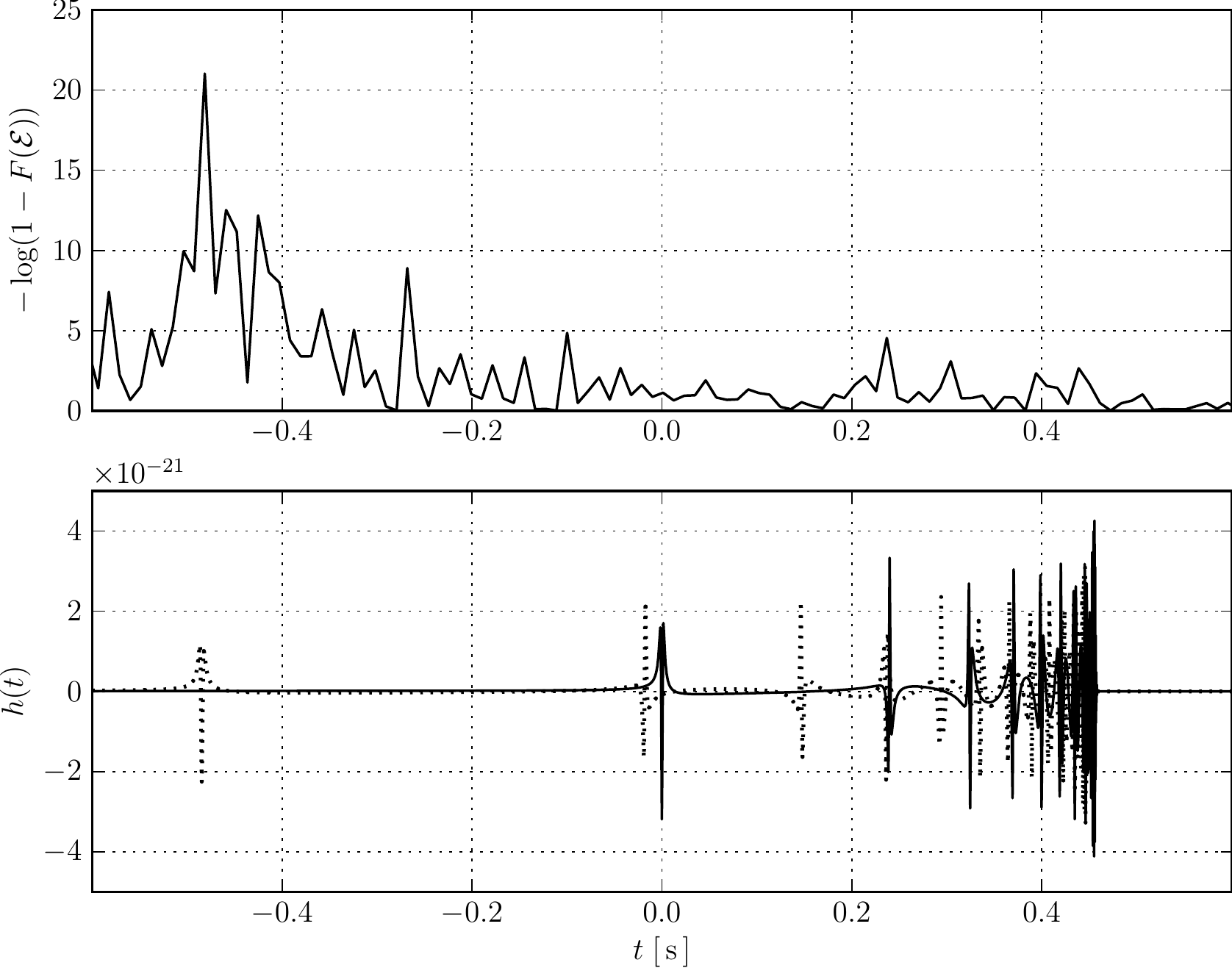}
\caption{The power stacking method in the mismatched case. The top panel shows the statistical significance of the computed statistic $\mathcal{E}$ plotted against time. In the bottom panel, the solid waveform shows the injected signal with binary parameters $M = 9\,M_{\astrosun}$, $r_{p} = 7.2\,M$, and $q=1$. The dotted waveform shows the $M=10\,M_{\astrosun}$, $r_{p} = 8\,M$, $q=1$ signal characterizing the signature at the time offset where the $\mathcal{E}$ statistic is maximal.}
\label{fig:qstack-example-mismatch}
\end{figure}

We now show example applications of the power stacking method in detecting a simulated signal injected into detector noise with a power spectrum given by the design requirements of Initial LIGO. We will consider two cases: (1) the (unrealistic) case where the signature parameters and actual signal parameters are identical, and (2) the case where there is some mismatch $\delta M$, $\delta r_{p}$, $\delta q$ in the parameters (in the following we will assume that the initial eccentricity $e_0=1$, such that the pair of compact objects are initially in an unbound parabolic trajectory).

In both examples, we use a signature generated for binary parameters (in geometric units) $M = 10\,M_{\astrosun}$, $r_{p} = 8\,M$, and $q=1$. After signature generation, we apply the stacking algorithm to realizations of Gaussian white noise in order to estimate a distribution of the $\mathcal{E}$ statistic with respect to this signature. The best-fit gamma distribution to this ensemble is shown in Fig.~\ref{fig:qstack-distribution}. The significance of any value of $\mathcal{E}$ is determined using the cumulative distribution function $F(\mathcal{E})$ of the best-fit gamma distribution. In the following, we express significance as the negative log probability of observing a value $\mathcal{E}' \, \geq \, \mathcal{E}$: $-\ln(1-F(\mathcal{E}))$. The threshold for an expected false detection rate of $1\,\mathrm{yr}^{-1}$ for a signature step size of $\delta t = 10\,\mathrm{ms}$ is $-\ln(1-F(\mathcal{E})) \approx 22$.

It is convenient to characterize the detectability of a signal by the optimal SNR obtained through matched filtering, $\rho_{\mathrm{opt}}$(see for example~\cite{Flanagan:1997sx}). In the following, we inject our simulated signals with $\rho_{\mathrm{opt}} = 14$, computed over the entire sequence of bursts. The output of the power stacking method in the matching case is shown in Fig.~\ref{fig:qstack-example-match}, while the output in the mismatched case is shown in Fig.~\ref{fig:qstack-example-mismatch}. The time axis in the top panels of Figs.~\ref{fig:qstack-example-match} and \ref{fig:qstack-example-mismatch} indicate the time offset of the signature; for example, in Fig.~\ref{fig:qstack-example-match}, the peak at $t = 0\,\mathrm{s}$ represents energy accumulated over the entire structure of the signal plotted in the bottom panel.

These examples only characterize single injections in noise and therefore are of limited use in characterizing the performance of the power stacking method. In the following section, we give the results of several tests of our method's performance.

\subsection{Results}
\label{sec:qstack-simulation}

We perform Monte Carlo simulations to characterize the performance of our power stacking algorithm. We also validate the robustness of the algorithm to parameter mismatch and modeling error. In these simulations, we run power stacking on ensembles of simulated signals injected into simulated noise with a power spectrum characteristic of the design sensitivities of Initial LIGO and Advanced LIGO. 

As this is a first test of performance, we did not attempt to model any non-stationarities, transients, or non-Gaussian characteristics of the simulated noise, though in practice these features are the primary source of false detections in single-detector searches (we do not consider multiple-detector searches here, which would significantly reduce false alarm rates). We note, however, that our method should be less susceptible to false detections due to transients in any single time-frequency tile, since we compute the total power over multiple tiles that are delocalized in time-frequency space.

\begin{figure}
\includegraphics[width=.45\textwidth]{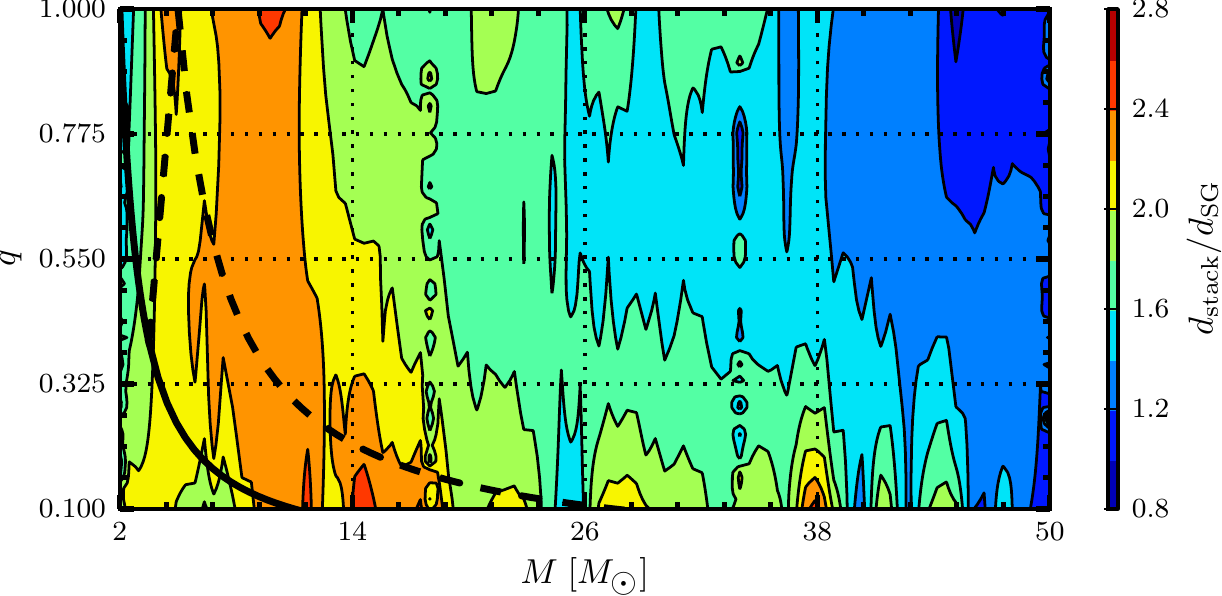}
\caption{Gain in horizon distance $d_{\mathrm{stack}}/d_{\mathrm{SG}}$ compared to a single-burst search using the Q-Pipeline for Advanced LIGO over a range of total system mass $M$ and mass ratio $q$ for fixed $r_{p} = 8M$. The top-left region corresponds to the space of NS-NS binaries; the middle region BH-NS binaries, and the rightmost region BH-BH binaries. In the following analysis, we focus on the mass range $2$--$25\,M_{\astrosun}$ where the method achieves the best performance.}
\label{fig:horizon-large}
\end{figure}

We focus on the range of initial periapse distances $r_{p}\, \leq \, 10M$. Larger initial periapse passages are certainly relevant for producing repeated burst phases that would be in the LIGO band (see the discussion in~\cite{east2013observing}), though here to illustrate we focus on the closer range which will produce the strongest emission in the high eccentricity phase. Also to keep the illustration simple we focus on NS and low to moderate stellar-mass BH systems, with the total mass in the range $2$--$25\,M_{\astrosun}$ and ratios $q$ ranging from $0.1$ to $1$. This choice of mass range is further motivated by the observation that our method achieves the greatest gains in this regime, as illustrated in Fig.~\ref{fig:horizon-large}.

\subsubsection{Horizon distance}
\label{sec:horizon-distance}

Any detection statistic will exhibit a negative correlation between its expected value and the distance to the source of the gravitational wave signal, by virtue of the fact that strain scales as $\mathrm{distance}^{-1}$. We define the horizon distance to be the luminosity distance between the source and the observer at which the expected value of the detection statistic equals the detection threshold. It is therefore a measure of the maximum distance out to which a source can be detected. 

We compare the horizon distances achieved by (i) the power stacking method, (ii) a Q-Pipeline search that identifies events by thesholding on the energy of single time-frequency tiles, and (iii) a search using an optimal filter. These horizon distances are computed for thresholds corresponding to false detection rates of $1\,\mathrm{yr}^{-1}$. For the optimal filter search (i.e. matched filtering using the optimal template), the cumulative distribution function (CDF) of matched filtering SNRs is $F(\rho) = \mathrm{erf}(\rho/\sqrt{2})$ \cite{cutler1993last}; for the Q-Pipeline, the CDF of tile energies is $F(Z) = 1- \exp(-Z)$ (see Sec.~\ref{sec:qtransform-statistics}); and for the power stacking search, the threshold is computed using best-fit distributions as described in Sec.~\ref{sec:qstack-statistics}. For a sampling rate equal to the LIGO sampling frequency of $16,384\,\mathrm{Hz}$, the thresholds for Q-transform normalized tile energies and optimal filtering are $Z = 27$ and $\rho = 7$ respectively.

\begin{figure}
\includegraphics[width=.45\textwidth]{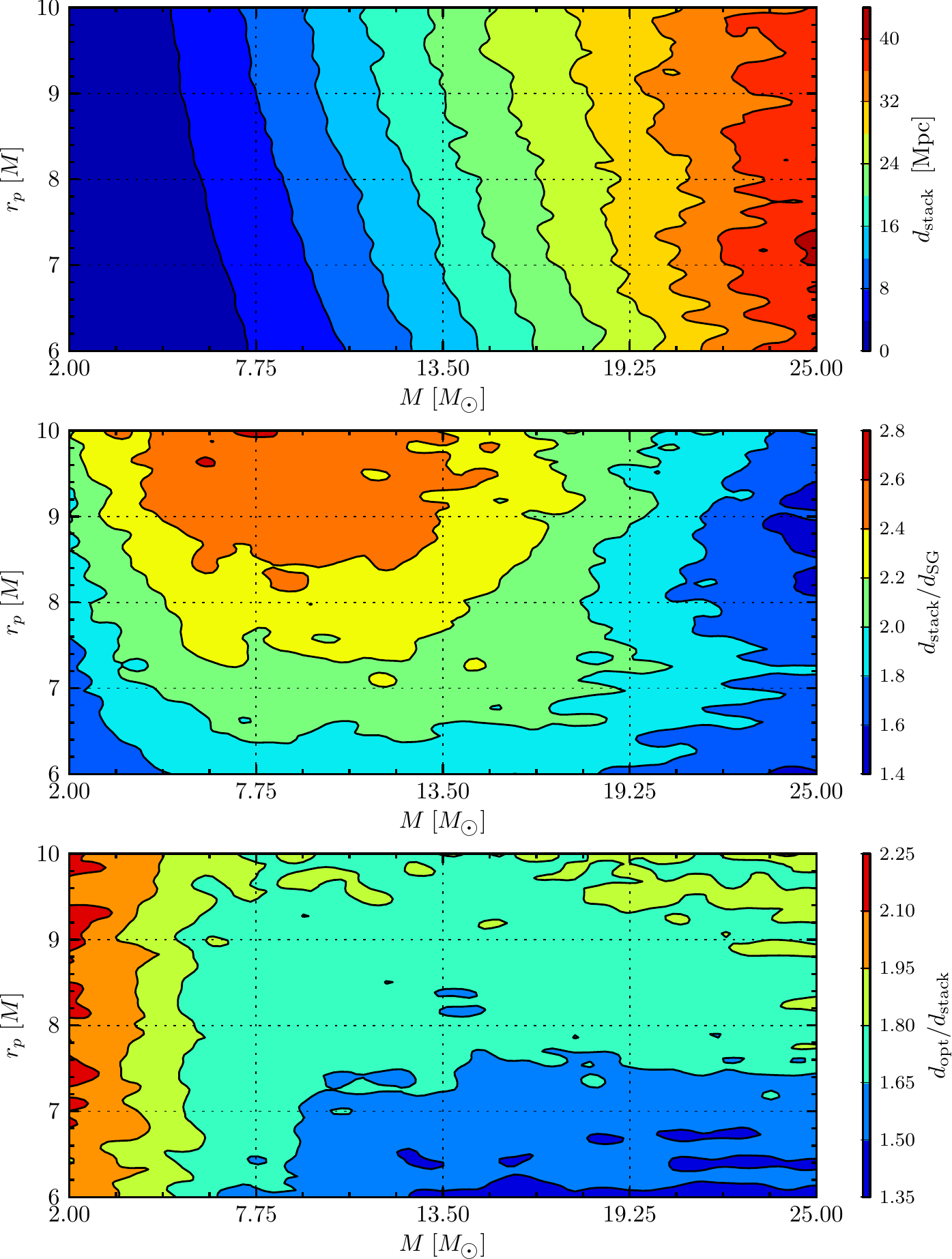}
\caption{Top: Contours of horizon distance for the power stacking method $d_{\mathrm{stack}}$ as a function of $M$ and $r_{p}$ for Initial LIGO and fixed $q = 1$. Middle: Gain in horizon distance $d_{\mathrm{stack}}/d_{\mathrm{SG}}$ compared to a single-burst search. Bottom: Loss in horizon distance $d_{\mathrm{opt}}/d_{\mathrm{stack}}$ compared to an optimal filter.}
\label{fig:horizon-mrp}
\end{figure}

\begin{figure}
\includegraphics[width=.45\textwidth]{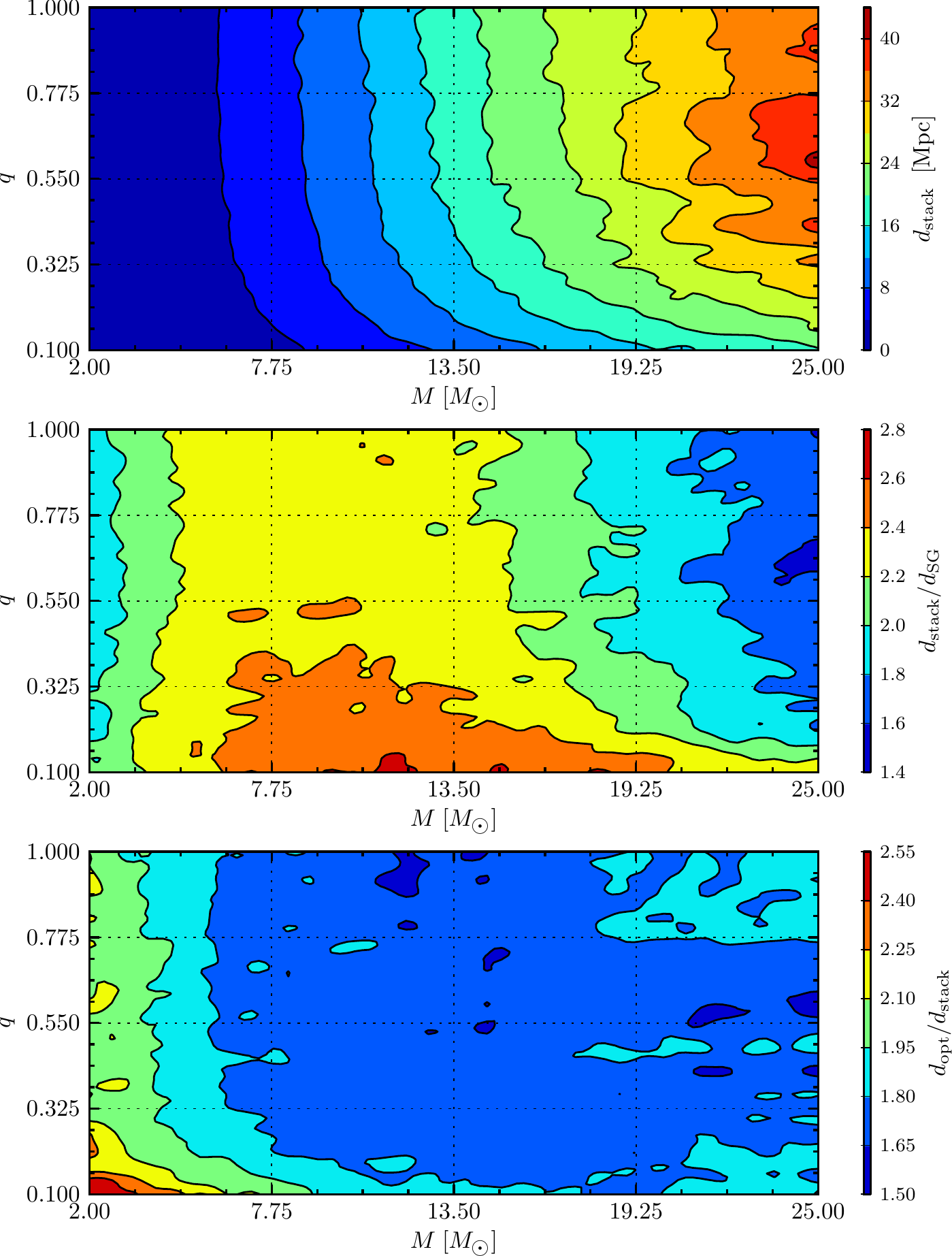}
\caption{Top: Contours of horizon distance for the power stacking method $d_{\mathrm{stack}}$ as a function of $M$ and $q$ for Initial LIGO and fixed $r_{p} = 8M$. Middle: Gain in horizon distance $d_{\mathrm{stack}}/d_{\mathrm{SG}}$ compared to a single-burst search. Bottom: Loss in horizon distance $d_{\mathrm{opt}}/d_{\mathrm{stack}}$ compared to an optimal filter.}
\label{fig:horizon-mratio}
\end{figure}

\begin{figure}
\includegraphics[width=.45\textwidth]{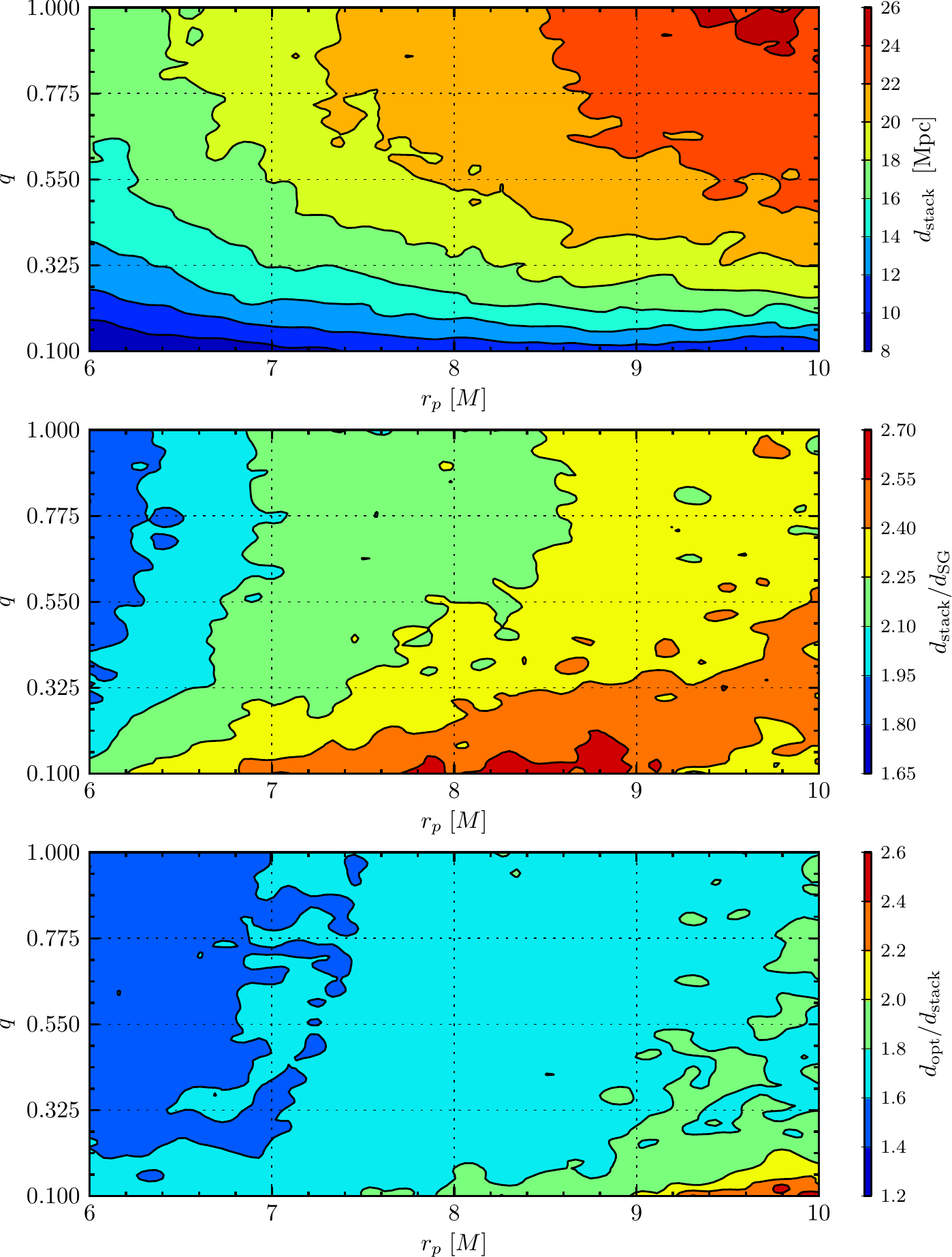}
\caption{Top: Contours of horizon distance for the power stacking method $d_{\mathrm{stack}}$ as a function of $r_{p}$ and $q$ for Initial LIGO and fixed $M=15\Msun$. Middle: Gain in horizon distance $d_{\mathrm{stack}}/d_{\mathrm{SG}}$ compared to a single-burst search. Bottom: Loss in horizon distance $d_{\mathrm{opt}}/d_{\mathrm{stack}}$ compared to an optimal filter.}
\label{fig:horizon-rpratio}
\end{figure}

The horizon distances are computed for a signal averaged over sky location, polarization angle and source orientations. Recall that the strain $h(t)$ observed by a gravitational wave detector is given by $h(t) = F_{+}(\theta, \phi, \psi)h_{+}(t) + F_{\times}(\theta, \phi, \psi)h_{\times}(t)$, where the detector response functions $F_{+}$ and $F_{\times}$ depend on the source sky position $(\theta, \phi)$ and the polarization angle $\psi$ between the two polarizations. The RMS average of gravitational wave strain over sky location and polarization angle of  $F_{+}$ and $F_{\times}$ is $\sqrt{\langle F^{2}_{+,\times} \rangle} = 1/\sqrt{5}$ \cite{thorne1987gravitational}. An optimally-oriented source would lie on the plane perpendicular to the line-of-sight from the Earth; to average over source orientations, we multiply the observed strain at a given distance by a factor of $\sqrt{\langle_{-2}Y_{2,\pm2} \rangle}=1/\sqrt{5}$, where $_{-2}Y_{m\ell}$ denotes spherical harmonics of spin weight $-2$. 

\begin{figure}
\includegraphics[width=.45\textwidth]{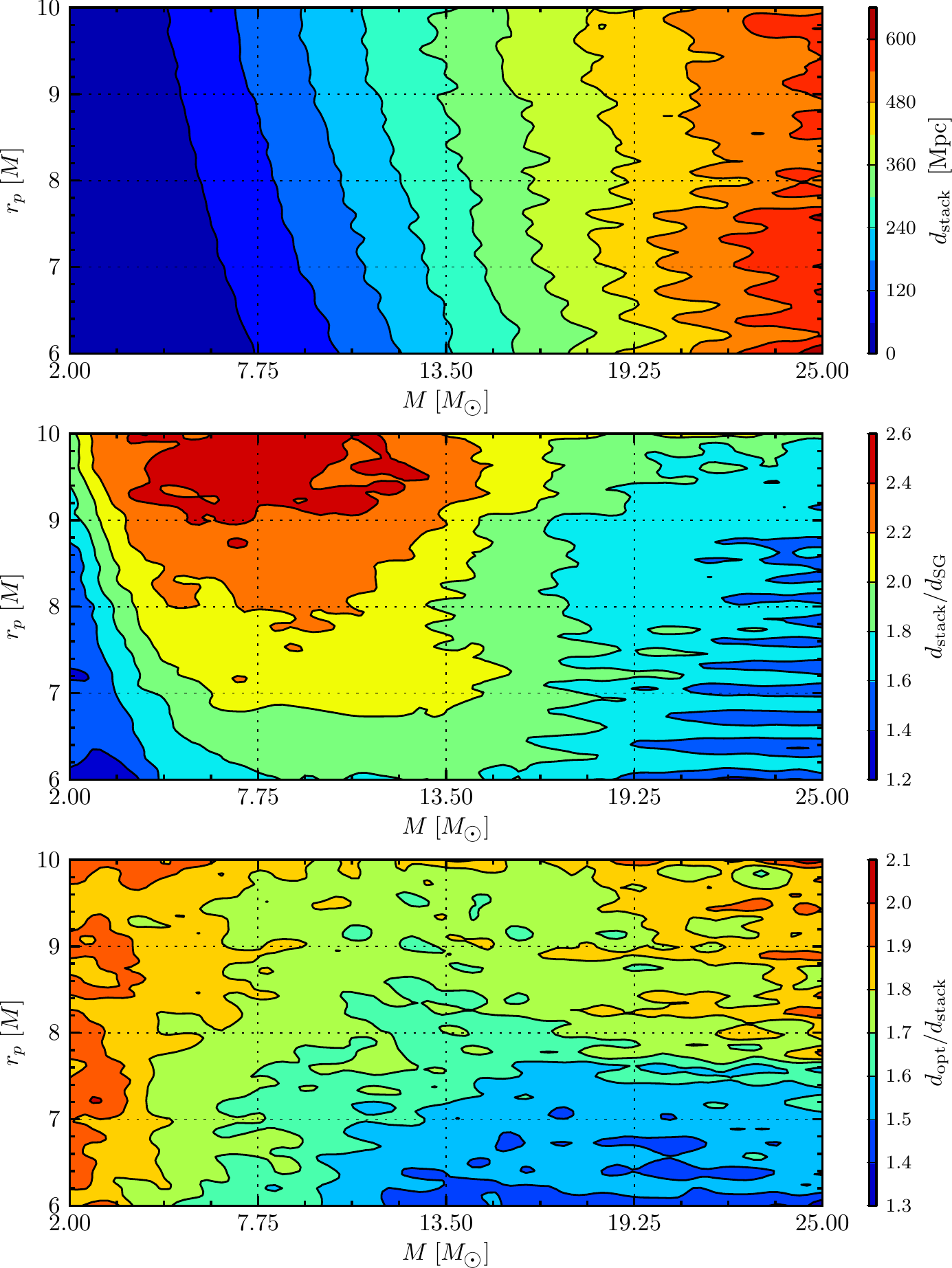}
\caption{Top: Contours of horizon distance for the power stacking method $d_{\mathrm{stack}}$ as a function of $M$ and $r_{p}$ for Advanced LIGO and fixed $q = 1$. Middle: Gain in horizon distance $d_{\mathrm{stack}}/d_{\mathrm{SG}}$ compared to a single-burst search. Bottom: Loss in horizon distance $d_{\mathrm{opt}}/d_{\mathrm{stack}}$ compared to an optimal filter.}
\label{fig:horizon-mrp-aligo}
\end{figure}

\begin{figure}
\includegraphics[width=.45\textwidth]{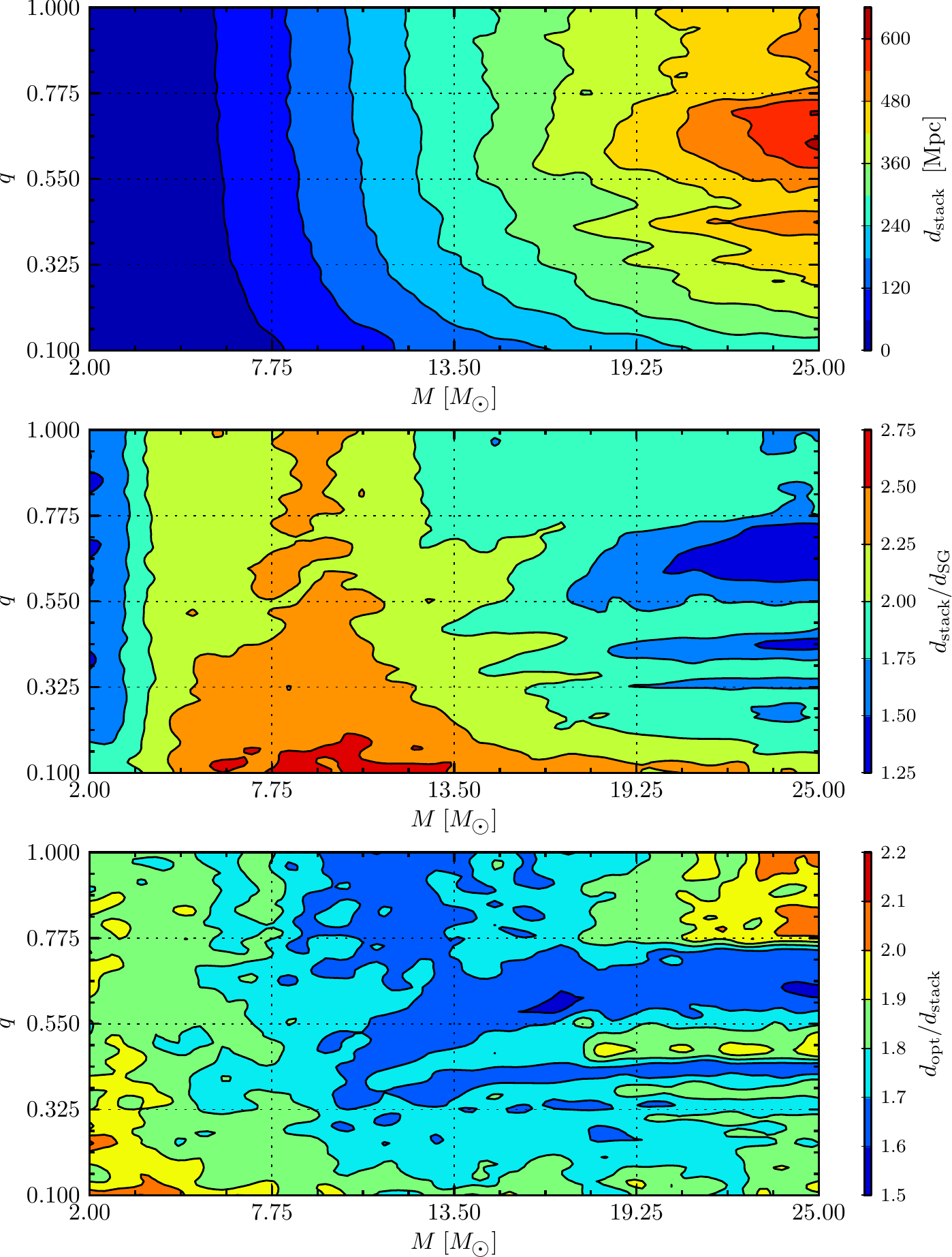}
\caption{Top: Contours of horizon distance for the power stacking method $d_{\mathrm{stack}}$ as a function of $M$ and $q$ for Advanced LIGO and fixed $r_{p} = 8M$. Middle: Gain in horizon distance $d_{\mathrm{stack}}/d_{\mathrm{SG}}$ compared to a single-burst search. Bottom: Loss in horizon distance $d_{\mathrm{opt}}/d_{\mathrm{stack}}$ compared to an optimal filter.}
\label{fig:horizon-mratio-aligo}
\end{figure}

\begin{figure}
\includegraphics[width=.45\textwidth]{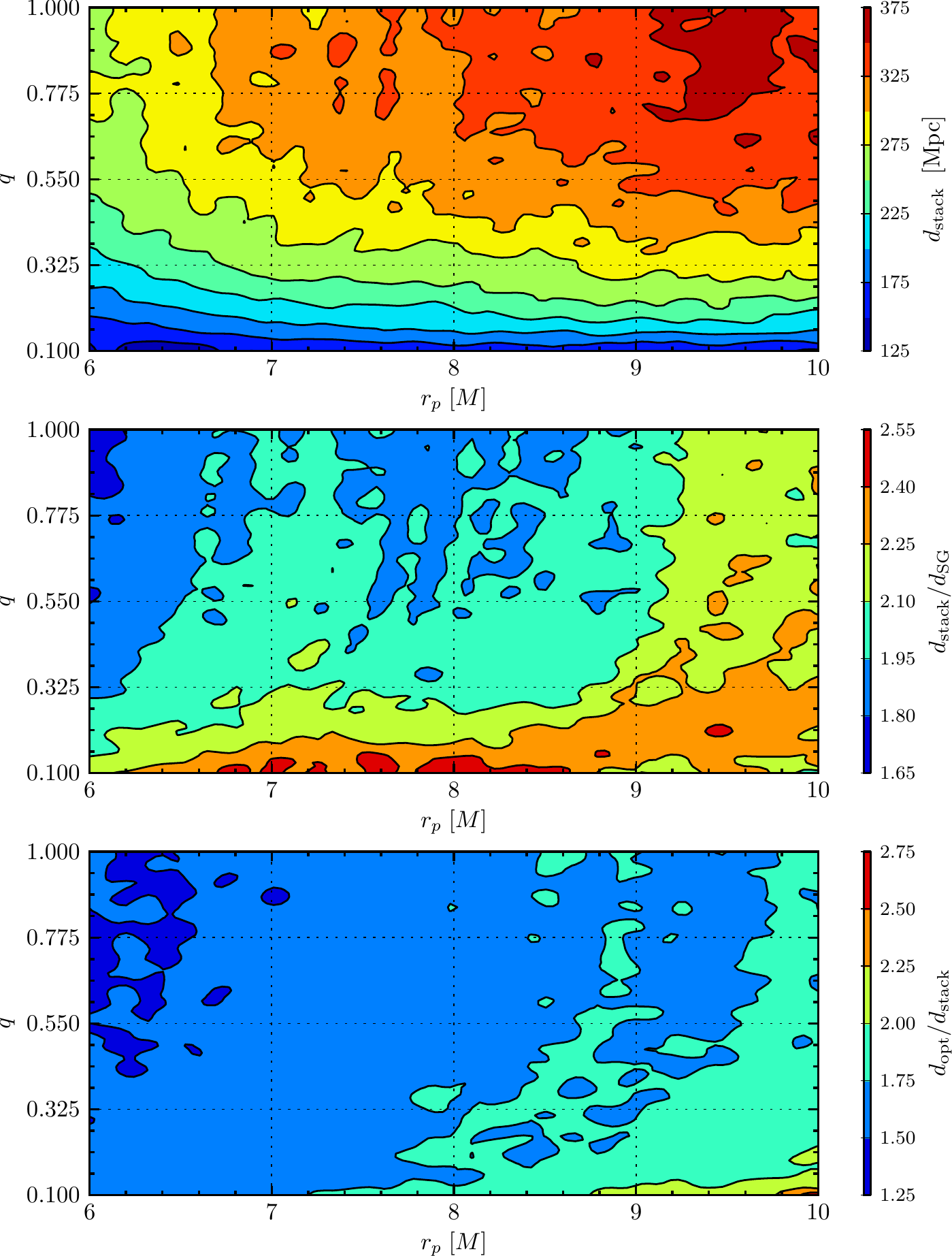}
\caption{Top: Contours of horizon distance for the power stacking method $d_{\mathrm{stack}}$ as a function of $r_{p}$ and $q$ for Advanced LIGO and fixed $M=15\Msun$. Middle: Gain in horizon distance $d_{\mathrm{stack}}/d_{\mathrm{SG}}$ compared to a single-burst search. Bottom: Loss in horizon distance $d_{\mathrm{opt}}/d_{\mathrm{stack}}$ compared to an optimal filter.}
\label{fig:horizon-rpratio-aligo}
\end{figure}

We generate contours of sky-averaged horizon distance for Initial LIGO (Figs.~\ref{fig:horizon-mrp}, \ref{fig:horizon-mratio}, and \ref{fig:horizon-rpratio}) and for Advanced LIGO (Figs.~\ref{fig:horizon-mrp-aligo}, \ref{fig:horizon-mratio-aligo}, and \ref{fig:horizon-rpratio-aligo}) (cf. the discussion above an optimally oriented source would be visible a factor of 5 times further away). The power stacking method improves on the method of identifying the single highest-energy time-frequency tile in the Q-transform by as much as a factor of $\sim$2-3 over much of parameter space (and event rate scales as the cube of this), with the largest gains observed for binaries with many repeating bursts (low $q$, high $r_{p}$). An optimal matched filter search offers a similar increase again in horizon distance over power stacking. In Figs.~\ref{fig:horizon-mrp} and \ref{fig:horizon-mrp-aligo}, we observe that power stacking achieves the greatest gain compared to the Q-Pipeline in the lower end of the mass range we consider. One can attribute this to the emphasis of the repeating burst phase by the detector sensitivity to the lower-mass regime, since the frequency of the signal varies inversely with the total mass of the binary. 

\subsubsection{Robustness to parameter mismatch}
\label{sec:robustness-to-parameter-mismatch}

\begin{figure}
\includegraphics[width=.45\textwidth]{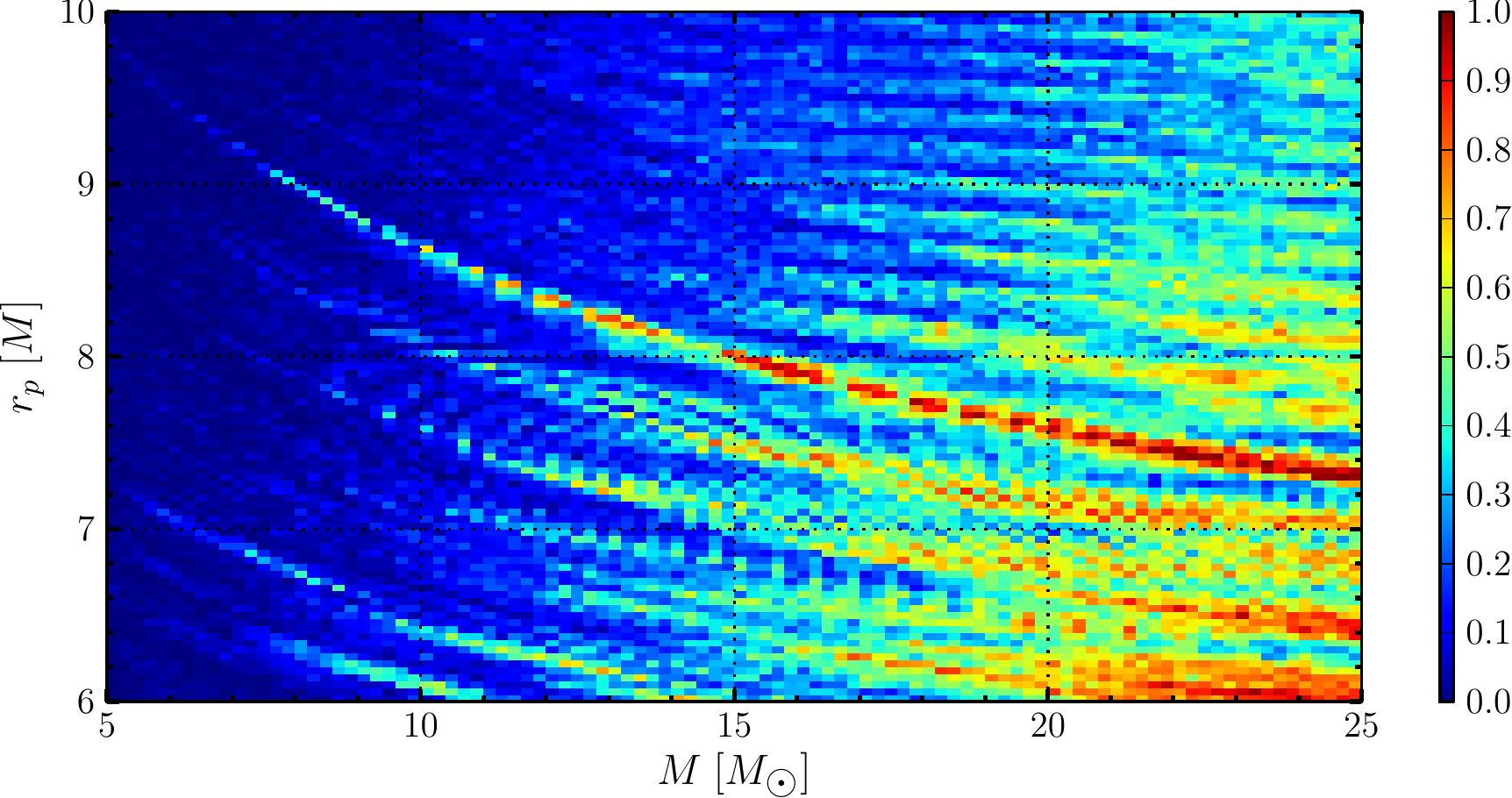}
\caption{Estimated detection probability over a range of $M$ and $r_{p}$ for a signature with fixed parameters $M=15\,M_{\astrosun}$, $r_{p} = 8\,M$, $q=1$ applied over a mismatched parameter space.
In this figure, and Figs.\ref{fig:param-mratio} and \ref{fig:param-rpratio}, we show the
raw data from the Monte Carlo studies vs. the contour plots of the corresponding data
in other plots; this is to more clearly illustrate the degeneracies in parameter space.}
\label{fig:param-mrp}
\end{figure}

\begin{figure}
\includegraphics[width=.45\textwidth]{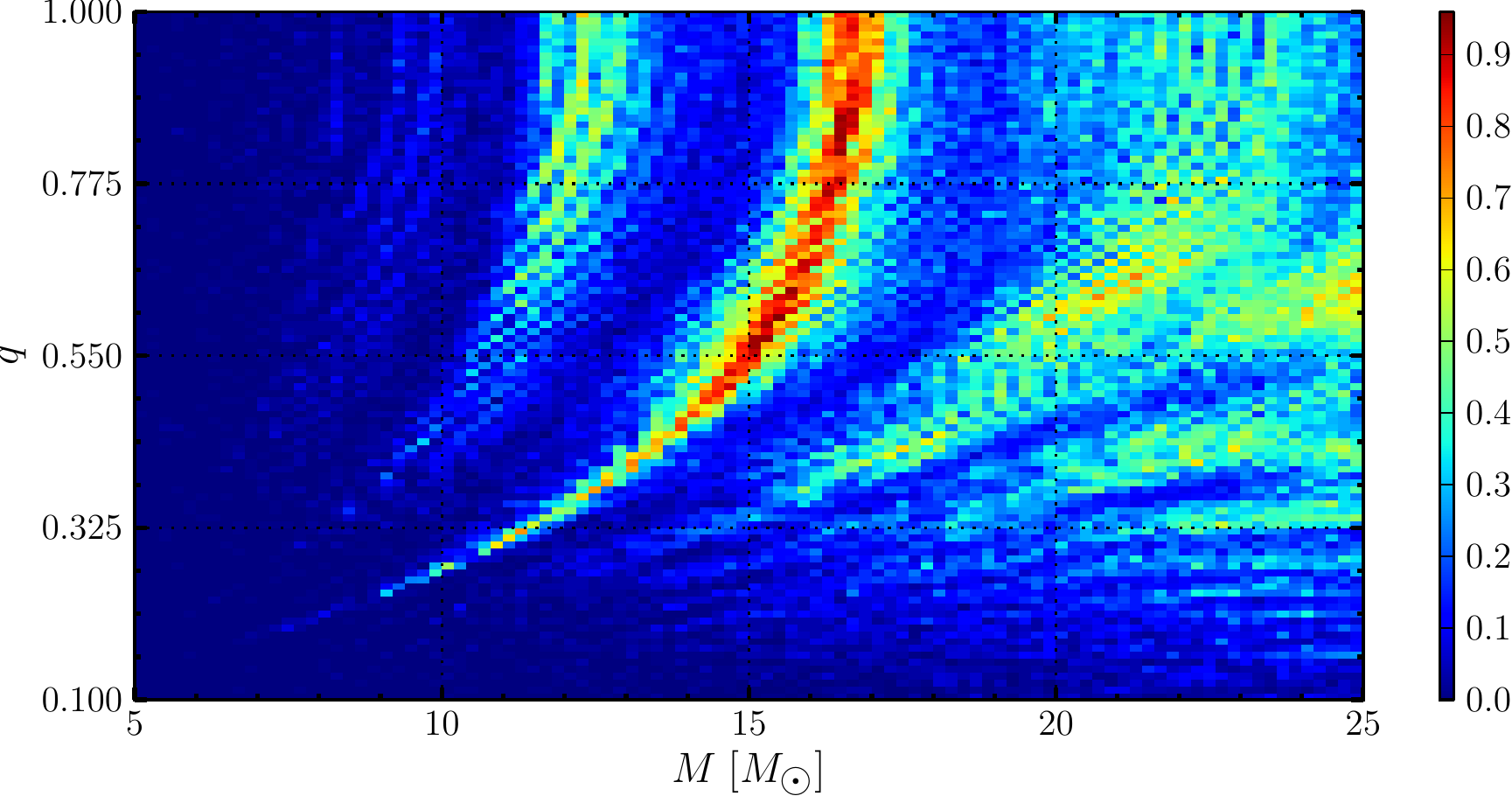}
\caption{Estimated detection probability over a range of $M$ and $q$ for a signature with fixed parameters $M=15\,M_{\astrosun}$, $r_{p} = 8\,M$, $q=0.55$ applied over a mismatched parameter space.}
\label{fig:param-mratio}
\end{figure}

\begin{figure}
\includegraphics[width=.45\textwidth]{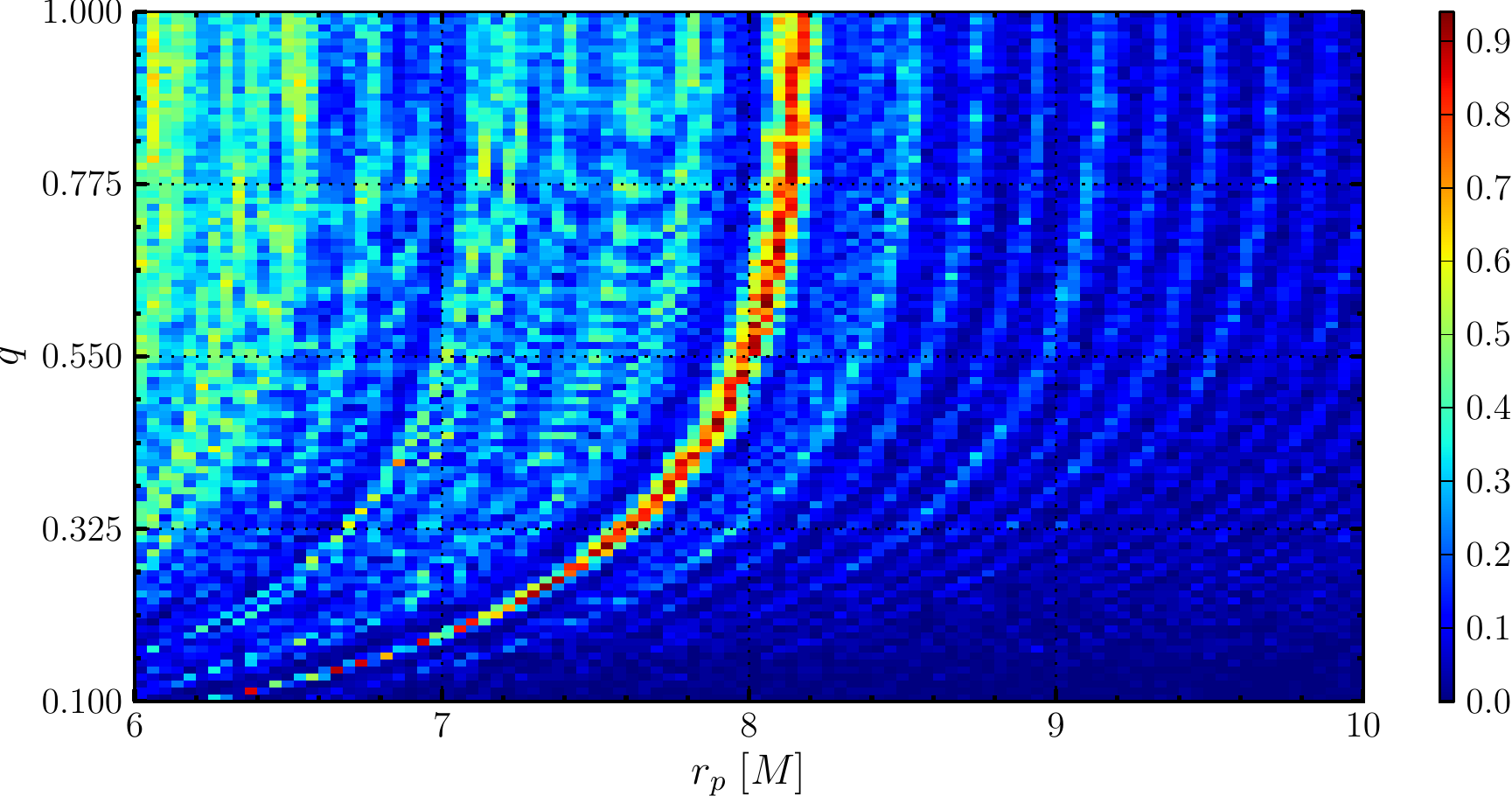}
\caption{Estimated detection probability over a range of $r_{p}$ and $q$ for a signature with fixed parameters $M=15\,M_{\astrosun}$, $r_{p} = 8\,M$, $q=0.55$ applied over a mismatched parameter space.}
\label{fig:param-rpratio}
\end{figure}

In a matched filter search, a template is only sensitive to signals in a very small region of parameter space around the signal that exactly matches the template (see ~\cite{Lindblom:2010mh,Damour:2010zb,Littenberg:2012uj} for some recent accuracy studies applied to quasi-circular inspirals). This implies one needs a rather dense covering of parameter space with templates, and more-over, as discussed in the introduction could require each template be computed with a level of accuracy difficult to achieve for eccentric binaries using existing numerical or perturbative methods. This motivated the development of the power stacking approach described here. Though we are not providing a comparison of accuracy requirements or template densities for matched filter vs. power stacking templates, in this and the following sub-sections we illustrate that (for detection) power stacking does not require an onorously dense template familiy and is rather robust to waveform modeling errors.

In Figs.~\ref{fig:param-mrp}, \ref{fig:param-mratio}, and \ref{fig:param-rpratio}, we show the variation of estimated detection probabilities when a fixed signature at the center of the parameter space is applied to signals with mismatched parameters. The optimal SNR of the signals is fixed at $\rho=14$. At each sample point in the parameter space, we inject the simulated signal into 100 noise realizations. We estimate the detection probability at each of these points as the fraction of instances where the maximum statistic exceeds the detection threshold.

We observe that the method achieves detection probabilities of at least $0.5$ over a wide range of parameters with a fixed signature. Therefore, these plots suggest that our parameter stacking method is robust to mismatch between the parameters of the signature and the parameters of the actual signal. This property is a result of the maximization over nearby basis functions around each best-match basis function $\psi_{k}$. We observe as well that for signals injected with fixed optimal SNR, the detection probability falls off quickly as $r_{p}$ increases. This is due to the spreading of the (fixed) signal energy over a large number of bursts in the inspiral phase. To account for this, we can consider a fixed value of $r_{p}$; in Fig.~\ref{fig:param-mratio}, we observe that for fixed $r_{p}$, the signature at the center of the parameter space achieves good detectability over a wide range of $M$ and $q$.

These results suggest that for purposes of signal detection (as opposed to parameter estimation or signal reconstruction), a large parameter space can be covered by a relatively sparse set of signatures, thereby reducing the computational cost of searching over the large parameter space of dynamical capture binaries. The curves of high detectability in Figs.~\ref{fig:param-mrp}, \ref{fig:param-mratio}, and \ref{fig:param-rpratio} are indicative of degenerate sets of signals over the parameter space. We expect the degenerate paths to follow contours of constant time intervals between successive periastra. Simple Keplerian dynamics give a reasonably good approximation for these paths; regions of parameter space that follow contours of constant $M(1+q)^2q^{-1}r_p^{3/2}$ are degenerate. This illustrates the trade-off between the size of the parameter space covered by each signature and the precision of source parameter estimation given a detection using a particular signature.

\subsubsection{Robustness to modeling error}

\begin{figure}
\includegraphics[width=.45\textwidth]{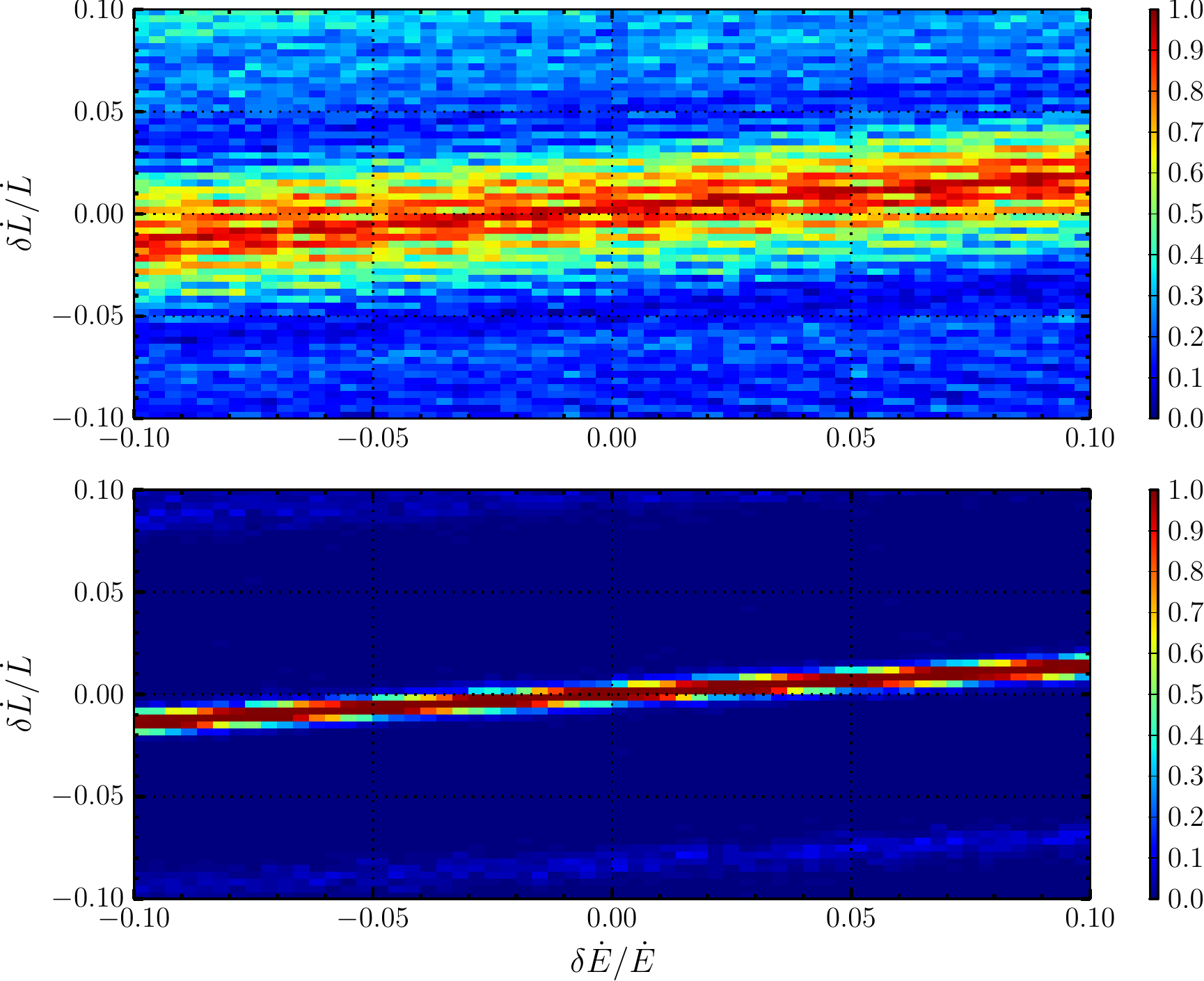}
\caption{Estimated detection probability as a function of modeling error in $\dot{E}$ and $\dot{L}$ for a signature with parameters $M = 15\,M_{\astrosun}$, $r_{p} = 8\,M$,  $q = 1$. Signals injected with $\rho = 14$. Top: Power stacking. Bottom: Matched filtering.}
\label{fig:model-error}
\end{figure}

To characterize the effect of modeling error, we treat the geodesic-based model discussed in Sec.~\ref{sec:model} as the correct signal, and simulate error terms by modifying the quadrupolar energy and angular momentum loss via:
\begin{align}
	\dot{E} &\rightarrow \dot{E} + \delta \dot{E}, \\
	\dot{L} &\rightarrow \dot{L} + \delta \dot{L},
\end{align} where $\delta \dot{E}$ and $\delta \dot{L}$ are constant multiples of $\dot{E}$ and $\dot{L}$ respectively. While crude, this error model serves to show some general properties of the algorithm's sensitivity to modeling error. A more thorough analysis would require a better understanding of the our model's actual error, which is beyond the scope of this work. 

Estimated detection probabilities as a function of $\delta \dot{E} / \dot{E}$ and $\delta \dot{L} / \dot{L}$ are plotted in Fig.~\ref{fig:model-error} for both the power stacking method and matched filtering. A fixed signature is used to search for signals injected with SNR $\rho=14$. As in Sec.~\ref{sec:robustness-to-parameter-mismatch}, the detection probability is estimated at each point using 100 simulations. By inspection, the detection probability is relatively insensitive to modeling error in $\dot{E}$ compared to $\dot{L}$. Note however that this is mostly an artifact of us modeling error as relative fractions of the respective luminosities in $E$ and $L$, and that quadrupole emission physics gives $\dot{L}/(\dot{E}M)\sim 1/(\omega M)\approx (r_p/M)^{3/2}$; hence for relevant values of $r_p$ a given fractional modification in $\dot{L}$ will have a larger effect on the resultant dynamics than the same fractional change in $\dot{E}$. This comparison demonstrates the increased robustness to modeling error relative to matched filtering that the power stacking method offers.

\section{Discussion and Conclusions}
\label{sec:discuss}

In this paper we have introduced a GW search strategy targeting highly eccentric binaries that, though sub-optimal compared to matched filtering, is more robust to modeling uncertainties, and promises improved sensitivity compared to existing unmodeled burst searches. The method is an adaptation of a power stacking algorithm developed in~\cite{kalmus2009stacking} to search for GW bursts associated with soft gamma ray repeater events. The primary difference in our method is the time-frequency signature over which power is summed is informed by a binary IMR model (the origin of gamma ray repeaters is unknown, so an unmodeled search makes sense for this putative source). In that regard it shares many of the issues and questions associated with template-based matched filter searches. These include the effects of modelling error on detection and parameter estimation, and how degeneracies in parameter space could limit uniquely identifying particular events. In this work we have illustrated some of these issues, though leave a more through examination to future work, as our main purpose here was to introduce the power stacking method for eccentric binaries. Within the framework described here there is much room for exploring variants of the basic method. We conclude with a discussion of possible future directions of study.

\subsection{Choice of metric and $\xi_{k}$ parameters}

Recall that for a given metric $D$, we consider the neighborhood $\{ \varphi \}$ around the basis function $\psi_{k}$ defined by $D(\varphi, \psi_{k})\, \leq \, \xi_{k}$. The $\xi_{k}$ parameters allow us to control how discriminating the search is. In the limit $\xi_{k}\rightarrow 0$, only the basis functions in the signature are considered, and consequently the method reduces to a variant of matched filtering with an approximate template. Larger values of $\xi_{k}$ increase robustness to parameter mismatch and modeling error, but at the cost of increasing the detection threshold and time complexity. The latter cost is due to the larger set of tiles that have to be considered at each time step, while the former is a statistical consequence of maximizing over a larger number of samples. It should therefore be possible to optimize the value of $\xi_{k}$ to obtain a desired balance between sensitivity (lower detection thresholds for a given false alarm rate) and robustness.

We also note that in the implementation described in Sec.~\ref{sec:qstack-implementation}, both the metric and $\xi_{k}$ parameters are chosen implicitly. The performance of other metrics is beyond the scope of this work, but possible candidates include, for example metrics that accept only basis functions with the same characteristic center frequency $\phi$:
\begin{equation}
	D(\psi_{1}, \psi_{2}) = 
	\begin{dcases}
		(\tau_{1} - \tau_{2})^{2} & \text{if }\phi_{1} = \phi_{2} \\
		\infty	& \text{otherwise}
	\end{dcases}
\end{equation}
Such a metric would be appropriate if it is known that the frequency of each burst is accurately modeled, but their relative timing is not.

\subsection{Choice of basis}

For reasons outlined in Sec.~\ref{sec:qstack-implementation}, we chose to implement our method using the Q-transform basis of multiresolution Gaussian-windowed exponentials. There exists a wide selection of other possible bases, most notably a multitude of wavelet families \cite{daubechies1990wavelet}. Briefly, a wavelet family is a set of functions $\{ \psi_{s,\tau}(t) \}$ defined by the translation and dilation of a single prototype $\psi(t)$, called the mother wavelet: 
\begin{equation}
	\psi_{s,\tau}(t) = \frac{1}{\sqrt{|s|}} \psi\left( \frac{t-\tau}{s} \right),
\end{equation}
where $s,\tau\in\mathbb{R}$ and $s\neq 0$. Like the Q-transform basis described in the preceding section, wavelet basis functions are localized in time and frequency, and the scale factor $s$ is a resolution parameter analogous to the $Q$ parameter. An orthonormal family of wavelets can be chosen such that the set forms a basis for the Hilbert space $L^{2}(\mathbb{R})$ of square-integrable functions. Wavelet transforms have been investigated as a basis for gravitational wave burst searches \cite{klimenko2004wavelet,klimenko2004performance,chatterji2004multiresolution,camarda2006search}.

Moreover, it is possible to design a mother wavelet $\psi(t)$ that matches a signal of interest such that the family $\{ 2^{-j/2} \psi(2^{-j}t-k) \}$, $j\in\mathbb{Z}$ is an orthonormal basis of $L^{2}(\mathbb{R})$ \cite{chapa2000algorithms}; thus, it should be possible to construct a wavelet transform that is adapted to the gravitational wave bursts of eccentric binaries. Such an adapted basis should increase the sensitivity of our method by more effectively concentrating the signal energy of each burst into a single time-frequency tile.

\subsection{Characterizing performance with realistic noise}

In Sec.~\ref{sec:qstack-simulation}, we performed a first characterization of the performance of our power stacking method using simulated detector noise. The simulated noise is stationary, which is not representative of actual detector noise. In searches for localized bursts of short duration ($<1\,\mathrm{s}$), it is usually assumed that the characteristics of detector noise are slowly-varying over the duration of a single burst, and therefore stationarity can effectively be assumed. However, these nonstationarities are more significant in a search for signals from eccentric binaries, since the repeating burst phase can include hundreds to thousands of bursts over minutes to days \cite{kocsis2012repeated}. The performance of our method with realistic noise therefore requires further study.

\subsection{Multiple detectors}

The method we described considers output from only a single detector. However, it is possible to extend the method to take advantage of networks of multiple gravitational wave detectors, such as the two 4 km LIGO detectors at Hanford and Livingston. Trivially, the search algorithm can be run on the output of each detector; a detection would then require coincident events in all outputs (within some time window determined by the light travel time between the detectors). A multiple detector search reduces the false alarm rate (and therefore increases sensitivity) since it is unlikely that a false detection due to noise transients would occur at multiple geographically separated detectors simultaneously. Additionally, a detection in a multiple detector search allows one to obtain an estimate of the sky-location of the source (see~\cite{Aasi:2013wya} for a review of the degree of localization possible with quasi-circular inpsirals).

\subsection{Enhanced parameter extraction with a hybrid search.}
As discussed in detail above, one of the primary reasons for introducing the power stacking method for eccentric binaries is to have a technique that does not need as high an accuracy for the model waveforms as matched filtering. This is because it is unlikely that computational resources will be available within the next few years to compute a template bank covering the full parameter space. The cost of power stacking is a reduction in detection rate compared to matched filtering. Furthermore, it is likely (though we have not investigated this here) that power stacking will likewise not be as effective as matched filtering in extraction of parameters from a putative detection. However, the parameter ranges inferred from a detection with a power stack searched may be sufficiently narrow that post-detection a high accuracy template bank within this range can be computed, and an improved estimation of the parameters made with a matched filter search. This would not mitigate the issue of lower detection rates with power stacking, though will allow maximal information about detected events to be gleaned.

\acknowledgments
This research was supported by NSF grants PHY-1065710, PHY-1305682 (FP) and the Simons Foundation (FP).  STM would like to thank the KITP at UCSB and the Aspen Center for Physics for their hospitality while parts of this work were completed.

% \clearpage

\bibliographystyle{prsty}

\bibliography{references}

\end{document}